\documentclass[submission, Phys]{SciPost}

\usepackage{empheq}
\newcommand{\bs}[1]{\boldsymbol{#1}}

\newcommand{\pa}{\partial}

\newcommand{\ket}{\rangle }
\newcommand{\bra}{\langle }

\def\widebar{\accentset{{\cc@style\underline{\mskip10mu}}}}

\begin{document}

\begin{center}{\Large \textbf{
Nonadiabatic Nonlinear Optics and Quantum Geometry  \\
--- Application to the Twisted Schwinger Effect 
}}\end{center}

\begin{center}
Shintaro Takayoshi\textsuperscript{1,2},
Jianda Wu\textsuperscript{3},
Takashi Oka\textsuperscript{1,4,5*}
\end{center}

\begin{center}
{\bf 1}
Max Planck Institute for the Physics of Complex Systems,
Dresden 01187, Germany
\\
{\bf 2} Department of Physics, Konan University,
Kobe 658-8501, Japan
\\
{\bf 3} Tsung-Dao Lee Institute \& School of Physics and Astronomy,
Shanghai Jiao Tong University, Shanghai 200240, China
\\
{\bf 4} Max Planck Institute for Chemical Physics of Solids,
Dresden 01187, Germany
\\
{\bf 5} The Institute for Solid State Physics, The University of Tokyo, Kashiwa, Chiba 277-8581, Japan
* oka@issp.u-tokyo.ac.jp
\end{center}

\begin{center}
\today
\end{center}


\section*{Abstract}
{\bf
We study the tunneling mechanism of nonlinear optical processes in 
solids induced by strong coherent laser fields. 
The theory is based on an extension of the 
Landau-Zener model with nonadiabatic geometric effects. 
In addition to the 
rectification effect known previously, we find two effects, namely 
perfect tunneling
and counterdiabaticity at fast sweep speed. 
We apply this theory to the 
twisted Schwinger effect, {\it i.e.},  nonadiabatic pair production of particles
by rotating electric fields, and
find a nonperturbative generation mechanism of the 
opto-valley polarization and photo-current
in Dirac and Weyl fermions. 
}

\vspace{10pt}
\noindent\rule{\textwidth}{1pt}
\tableofcontents\thispagestyle{fancy}
\noindent\rule{\textwidth}{1pt}
\vspace{10pt}

\section{Introduction}
\label{sec:intro}

Today, geometric effects~\cite{Berry1984} in electron dynamics
have become a central research topic 
in condensed matter~\cite{Xiao2010RMP}. 
In adiabatic processes, it is known that 
electrons acquiring a geometric phase provoke exotic effects 
such as quantum Hall effect~\cite{Thouless1982PRL,Kohmoto1985AnnPhys}. 
On the other hand, the importance of geometric effects 
in nonadiabatic processes have been overlooked 
except for a few examples such as the geometric amplitude 
factor~\cite{Berry1990ProcRoy,Nakamura1994PRA,Bouwmeestert1996JModOpt} and 
counterdiabatic driving~\cite{Demirplak2003JPCA,Berry2009JPA,delCampo2013PRL}
as well as the modification of the adiabaticity condition\cite{Wu2008PRA,Xu2018PRA}. 
In an example of nonadiabatic dynamics governed by a time-dependent Hamiltonian,
M.V. Berry showed that the tunneling probability 
can depend on the direction of the parameter sweep 
due to the geometric amplitude factor~\cite{Berry1990ProcRoy}.

We revisit the problem of nonadiabatic geometric effects
with a motivation to apply it to the 
twisted Schwinger effect in Dirac and Weyl Fermions. 
The Schwinger effect is 
fermion-antifermion pair production 
in strong electric fields~\cite{Schwinger1951PR,Sauter1931ZPhys,Heisenberg1936ZPhys}
and is known to originate from nonadiabatic 
tunneling in the 
momentum space~\cite{Zener1934ProcRoy,Kane1960JPCS,Oka2003PRL,OkaAoki2005PRL,Oka2012PRB}. 
Previously, AC extensions of the Schwinger effect were studied 
for linearly polarized fields 
$E_{x}=E\cos(\Omega t)$~\cite{Brezen1970PRD,Popov1974SovJNP,SchutzholdPRL2008}.  
The results have common nature as the problems of 
strong-field ionization~\cite{Krausz2009RMP} and a
particle escaping from an oscillating trap~\cite{Keldysh1965JETP}.
For low frequency, the tunneling  is 
exponentially suppressed with a threshold known
as the Schwinger limit~\cite{Schwinger1951PR,Sauter1931ZPhys,Heisenberg1936ZPhys}. 
For higher frequency (but still lower than the excitation gap), 
multiphoton excitation is activated and the excitation probability obeys a power law. 
Nonadiabatic geometric effects 
kicks in when we study 
pair production induced by rotating electric fields (or circularly polarized laser fields) 
$E_{x}+iE_{y}=Ee^{i\Omega t}$ \cite{PhysRevD.89.085001,kelardeh2021photoinduced},
which we coin as the ``twisted Schwinger effect''. 
If we assume momentum conservation, the problem of the twisted Schwinger effect 
can be recast to the Landau-Zener problem with a curved trajectory in the parameter space. 
This effective model is reminiscent of the twisted Landau-Zener model 
studied by M.V. Berry mentioned above~\cite{Berry1990ProcRoy}.
We perform a numerical analysis of the effective model dynamics
and find three geometric effects.
The first is the sweep direction dependence, 
which we call rectification. This is the same phenomenon indicated in ref.~\cite{Berry1990ProcRoy}
in terms of the geometric amplitude factor. The two other effects are perfect tunneling
and counterdiabaticity at fast sweep speed. 
In order to clarify the origin of the effects, 
we ``untwist'' the model with a unitary transformation, and obtain the standard Landau-Zener model
with an effective gap parameter depending on the geometric amplitude factor (see Eq.\eqref{eq:effectivegap} below).  
We can understand the three nonadiabatic geometric effects in a unified way
through the modulation of the effective gap. 
Recently, 
rectification in quantum tunneling has been studied in solid-state systems~\cite{Kitamura2020CommPhys}. 
However, as far as we know, the perfect tunneling
and counterdiabaticity at fast sweep has not been argued in previous studies.

In a condensed matter framework, a rotating electric field 
is created by a circularly polarized laser~\cite{Karch2010PRL,Wang2013Sci,Mciver2020NatPhys}, 
or shaking an optical lattice~\cite{Jotzu2014Nat},
while in high energy physics, it mimics the field 
created by ions passing by 
each other in heavy-ion collision experiments~\cite{Voronyuk2011PRC}. 
The rotating electric fields 
are known to induce valley polarization~\cite{Yao2008PRB,Xiao2012PRL} 
and photo-currents in 2D and 3D Dirac/Weyl materials~\cite{HosurPRB2011,JuanNatCom2017,Chan2017PRB}, respectively. 
Second order perturbation~\cite{SipePRB2000}
has served as a theoretical framework to describe these phenomena. 
Due to the development of strong coherent laser sources, an extension of the theory 
to the nonperturbative regime is being awaited.  
We show that the three nonadiabatic geometric effects, i.e., rectification, perfect tunneling
and counterdiabaticity, play an important role in 
understanding the nonperturbative versions of the opto-valley polarization
and photo-currents in 2D and 3D Dirac/Weyl materials 
which are microscopically caused by the twisted Schwinger effect. 
On the other hand, if these symmetries 
are broken, it is possible to realize finite $U(1)$ photocurrent in 
a similar way as in the optical absorption mechanism 
proposed in \cite{Chan2017PRB,Ma2017NatPhys,Chang2020PRL}.


\section{Nonadiabatic geometric effects in quantum tunneling}

\begin{figure}[htb]
\centering
\includegraphics[width=0.92\textwidth]{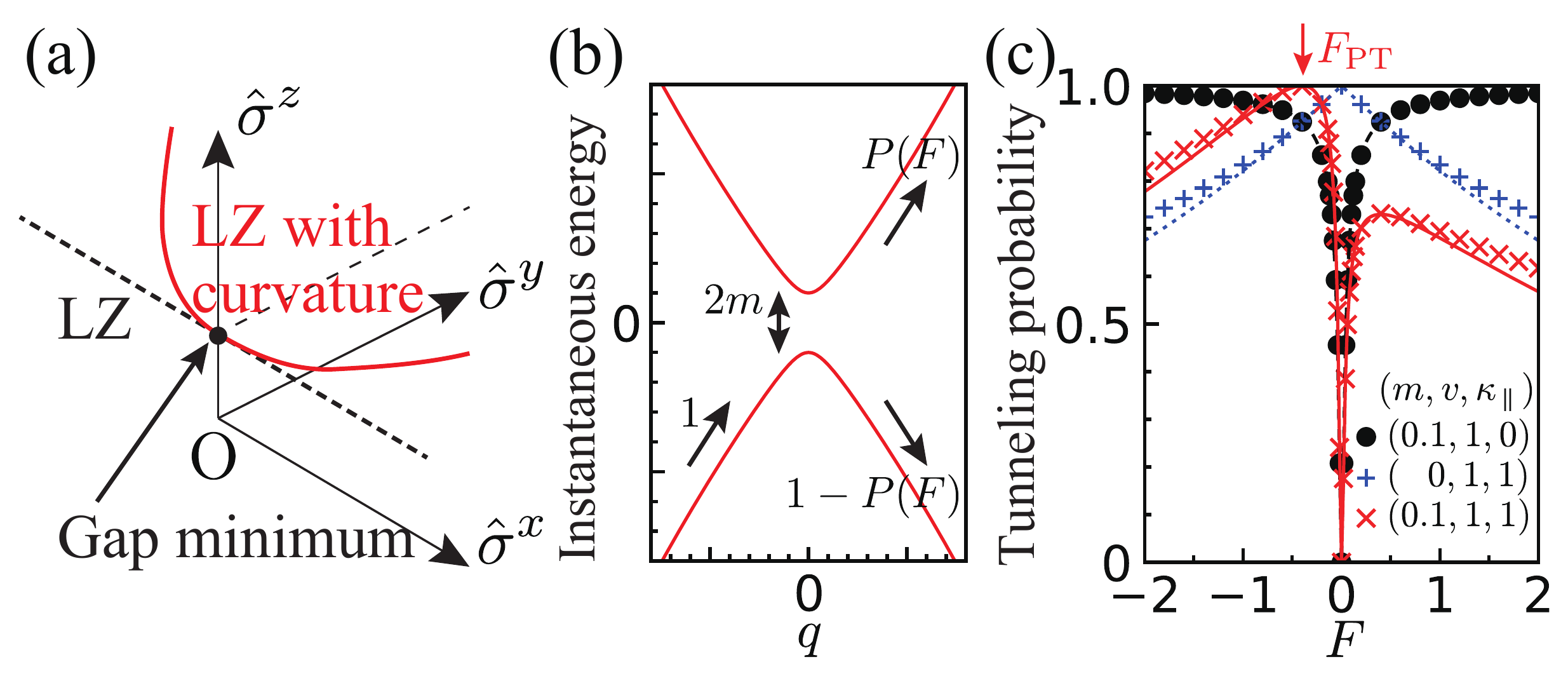}
\caption{
\textbf{Nonadiabatic geometric effects}:
(a) Schematic picture of the LZ tunneling with 
curvature in parameter space. 
(b) Instantaneous energy of the Hamiltonian Eq.~\eqref{eq:Hamil_q2} 
with $(m,v,\kappa_\parallel)=(0.1,1,1)$ 
and schematic picture for quantum tunneling. 
(c) Tunneling probability $P(F)$ for Eq.~\eqref{eq:Hamil_q2} 
with a parameter sweep $q=-Ft$ obtained numerically (marks)
compared with the tunneling formula Eq.~\eqref{eq:LZquad} (lines). 
}
\label{fig:Minimal}
\end{figure}

We demonstrate the nonadiabatic geometric effects 
in a two level Hamiltonian with a parameter $q$ defined by
\begin{align}
 \hat{\mathcal{H}}(q)=m\hat{\sigma}^{z}+vq\hat{\sigma}^{x}
   +\frac{1}{2}\kappa_\parallel v^{2}q^{2}\hat{\sigma}^{y},
\label{eq:Hamil_q2}
\end{align}
where $\hat{\sigma}^{j}$ ($j=x,y,z$) is the Pauli matrix, 
$m$ is the gap, and $v\;(>0)$ the energy slope. 
We use the unit $\hbar=c=1$. 
The model Eq.~\eqref{eq:Hamil_q2} is a quadratic expansion of 
the twisted LZ model introduced by M. V. Berry~\cite{Berry1990ProcRoy}. 
We consider the diabatic tunneling problem in this Hamiltonian $\hat{\mathcal{H}}(q)$ which is formulated as follows. 
\begin{enumerate}
\item We select an initial time $t=t_i (<0)$. This is selected to be far enough from the 
time $t=0$ when anti-crossing occurs. 
\item Between the two eigenstates $ \hat{\mathcal{H}}(q_i)|\pm(q_i)\ket=E_\pm(q_i)|\pm(q_i)\ket$ with $q_i=q(t_i)$,
we select the initial state to be the lower energy eigenstate $|-(q_i)\ket$. 
\item The state evolves from the initial time $t=t_i$ to the final time
$t=t_f\equiv |t_i|$ according to the Hamiltonian $\hat{\mathcal{H}}(q(t))$. 
\item The solution is given as $|\psi(t_f)\ket=\alpha |-(q_f)\ket+\beta |+(q_f)\ket$
where $|\pm (q_f)\ket$ are the two eigenstates of $\hat{\mathcal{H}}(q_f)$ with $q_f=q(t_f)$. 
The tunneling probability is given by $P(F)=|\beta|^2$. 
\end{enumerate}

If we regard the coefficients of the Pauli matrices 
$\boldsymbol{x}(q)=(vq,\frac{1}{2}\kappa_\parallel v^{2}q^{2},m)$ 
as a trajectory in the three-dimensional (3D) space,
it defines a curve and $\kappa_{\parallel}$ is the geodesic curvature 
around the gap minimum in the parameter space [Fig.~\ref{fig:Minimal}(a)].
The case of $\kappa_{\parallel}=0$ corresponds to 
the Landau-Zener (LZ) Hamiltonian~\cite{Landau1932ZSow,Zener1934ProcRoy}. 
The instantaneous energy of this Hamiltonian is plotted in Fig.~\ref{fig:Minimal}(b).
The tunneling probability $P(F)$ for a linear parameter sweep $q=-Ft$ 
in Eq.~\eqref{eq:Hamil_q2} can be evaluated and becomes (see subsection~\ref{sec:derivation} for derivation)
\begin{align}
 P(F)=\exp\bigg[
   -\pi
     \frac{(m+\kappa_{\parallel}vF/4)^{2}}{ v|F|}\bigg].
\label{eq:LZquad}
\end{align}
Comparing this expression with the LZ formula, 
we notice that the effective tunneling gap is modified by the geodesic curvature.

The nonadiabatic geometric effects in the tunneling probability Eq.(\refeq{eq:LZquad})
can be related to the Berry connection and quantum geometry. 
Using the instantaneous eigenstates of the Hamiltonian satisfying
\begin{align}
\hat{\mathcal{H}}(q)|\psi_m(q)\rangle
   =E_m(q)|\psi_m(q)\rangle,
\end{align}
with $m=\pm$, we define the Berry connection 
\begin{align}
\mathcal{A}_{nm}(q)=\langle\psi_{n}(t)|i\partial_{q}|\psi_{m}(t)\rangle. 
\end{align}
The Berry connection relates the basis sets $|\psi_m(q)\rangle$
spanned by the instantaneous eigenstates
at different parameter points $q$. 
We can define a gauge independent quantity 
\begin{align}
R_{nm}(q)=-A_{nn}(q)+A_{mm}(q)+\partial_{q}\arg A_{nm}(q)
\end{align}
known as the geometric amplitude factor~\cite{Berry1990ProcRoy} or the 
quantum geometric potential~\cite{Wu2008PRA,Xu2018PRA}. 
In the  Berry phase theory 
of polarization~\cite{Resta1994RMP}, where $q$ is regarded as the momentum in solids,
$R_{nm}(q)$ is known as the shift vector 
that corresponds to the 
difference of the electric polarization between the $n$ and $m$-th bands~\cite{SipePRB2000}. 
In particular,

Quantum tunneling in the presence of the geometric amplitude factor has been studied~\cite{Berry1990ProcRoy,Wu2008PRA,Xu2018PRA,Kitamura2020CommPhys}
and it was pointed out that this factor strongly affects 
the adiabaticity condition~\cite{Wu2008PRA,Xu2018PRA}. 
We can see this by rewriting the tunneling probability using the geometric amplitude factor. 
The parameter $\kappa_\parallel$ 
 in our quadratic Hamiltonian \eqref{eq:Hamil_q2}
is related to the geometric amplitude factor by 
\begin{align}
R_{+-}(q=0)=v\kappa_\parallel,
\end{align}
and we can write the tunneling probability as ($\Delta =E_+-E_-$ at $q=0$)
\begin{align}
 P(F)
   =&\exp\bigg[
      -\frac{\pi}{4v|F|}
        \left(\Delta +\frac{FR_{+-}}{2}\right)^{2}\bigg].
\label{eq:LZquadSM2}
\end{align}
This expression shows how quantum geometry affects the nonadiabatic tunneling process, 
where the effective tunneling gap is modified to 
\begin{align}
 \Delta_{\rm eff}=\Delta +\frac{FR_{+-}}{2}.
\label{eq:effectivegap}
\end{align}
This expression reproduces the generalized adiabaticity condition obtained by one of the present authors in Refs.~\cite{Wu2008PRA,Xu2018PRA}.

The tunneling formula \eqref{eq:LZquadSM2} predicts several interesting phenomena as we list below.  
\begin{description}
\item[ Rectification] 
Although the instantaneous band structure is symmetric in 
$q\to -q$, 
the tunneling probability depends on the sign of $F$
and rectification happens~\cite{Berry1990ProcRoy}. 
The ratio 
$\gamma(F)\equiv P(|F|)/P(-|F|)=\exp\left(-\pi \frac{ \Delta  R_{+-}}{2v}\right) $
deviates from unity for $m\neq 0$ 
[Fig.~\ref{fig:Minimal}(c)].

\item[Perfect tunneling]
In conventional LZ tunneling, the tunneling probability 
monotonically increase from 0 (adiabatic) to 1 (diabatic limit or perfect tunneling)
as the sweep speed increase.   
However, in the presence of nonadiabatic geometric effects,
perfect tunneling is realized at finite sweep speed. 
For $m\neq 0$, $P(F)$ peaks out and becomes unity 
at a perfect tunneling sweeping speed $F_{\mathrm{PT}}=-2\Delta/R_{+-}$ 
indicated by an arrow in Fig.~\ref{fig:Minimal}(c),
which is determined from the condition  $\Delta_{\mathrm{eff}}=0$.

\item[Counterdiabaticity at fast sweep]
For large $|F|$, 
$P(F)$ decreases as 
$\exp(-\pi R_{+-}^2|F|/16v)$. 
In the extreme case of $m=0$, 
the tunneling probability is a monotonically 
decreasing function of speed. 
\end{description}

We have performed a numerical calculation of the tunneling probability 
using the Hamiltonian Eq.~\eqref{eq:Hamil_q2}
and compared it with the tunneling formula  Eq.~\eqref{eq:LZquad} as depicted in Fig.~\ref{fig:Minimal}(c). 
The results show good agreement and the above three 
nonadiabatic geometric effect is clearly seen.

For convenience, we also consider
the two-band Hamiltonian with general operators up to $q^{2}$ order,
\begin{align}
 \mathcal{H}=\hat{A}+\hat{B}q+\hat{C}q^{2}/2.
\label{eq:HamilGeneralOpSM}
\end{align}
The gap minimum and velocity extremum conditions at $q=0$ require
$\{\hat{A},\hat{B}\}=0$ and $\{\hat{B},\hat{C}\}=0$, respectively.
This Hamiltonian is equivalent to the case of Eq.~\eqref{eq:Hamil_q2_SM} 
with the parameters 
\begin{align}
 m=\|\hat{A}\|,\quad
 v=\|\hat{B}\|,\quad
 \kappa_{\parallel}v^{2}=-\frac{i}{8}
   \frac{\mathrm{Tr}\{[\hat{A},\hat{B}],\hat{C}\}}{\|\hat{A}\|\|\hat{B}\|},
\label{eq:ABC}
\end{align}
where 
$\|\hat{O}\|\equiv\frac{1}{2}\sqrt{\mathrm{Tr}\{\hat{O},\hat{O}\}}$.

\subsection{A detailed derivation of the tunneling formula}
\label{sec:derivation}
In this subsection, we explain the derivation of the tunneling formula 
(Eq.~\eqref{eq:LZquad}) for the Hamiltonian 
\begin{align}
 \hat{\mathcal{H}}(q)=m\hat{\sigma}^{z}+vq\hat{\sigma}^{x}
   +\frac{1}{2}\kappa_\parallel v^{2}q^{2}\hat{\sigma}^{y},
\label{eq:Hamil_q2_SM}
\end{align}
where $\hat{\sigma}^{j}$ ($j=x,y,z$) is the Pauli matrices, 
$m$ is the gap, $v$ the energy slope, 
and $\kappa_{\parallel}$ is the curvature around the gap minimum 
in the parameter space. 
The idea is to move to a local frame with trivial geometry, 
which we call the ``LZ frame'', 
and use the LZ formula or its extension: 
the Dykhne-Davis-Pechukas (DDP) 
(also known as the Landau-Dykhne or the imaginary time) 
method~\cite{Dykhne1962JETP,Davis1976JCP} 
(see Ref.~\cite{Oka2012PRB} for an extended discussion of the method). 

Let us start from a general two-band Hamiltonian 
\begin{align}
 \hat{\mathcal{H}}(q)
   =\boldsymbol{d}(q)\cdot\hat{\boldsymbol{\sigma}},
 \label{eq:generalH_SM}
\end{align}
where $\boldsymbol{d}(q)$ defines a curve 
in the Euclidean space. 
We consider tunneling at the gap minimum $q=0$, 
and define the unit directional, tangential, and normal vectors as 
\begin{align}
 \boldsymbol{r}=&
   \boldsymbol{d}(0)/|\boldsymbol{d}(0)|\nonumber\\
 \boldsymbol{t}=&
   \partial_{q}\boldsymbol{d}(0)/|\partial_{q}\boldsymbol{d}(0)|\nonumber\\
 \boldsymbol{n}=&
   \boldsymbol{r}\times\boldsymbol{t}.\nonumber
\end{align}
Note that $\boldsymbol{t}\perp\boldsymbol{r}$. 
We move to the LZ frame, 
where the curve $\boldsymbol{d}(q)$ is transformed to 
a curve on the plane spanned by $\boldsymbol{r}$ and $\boldsymbol{t}$ 
using a unitary operator 
$\hat{U}=e^{i\frac{\theta(q)}{2}\boldsymbol{r}\cdot\hat{\boldsymbol{\sigma}}}$. 
The angle $\theta(q)$ is determined as 
\begin{align}
 \hat{U}^{\dagger}\hat{\mathcal{H}}(q)\hat{U}
   =[a(q)\boldsymbol{r}+b(q)\boldsymbol{t}]
     \cdot\hat{\boldsymbol{\sigma}},\nonumber
\end{align}
where 
$a(q)=\boldsymbol{d}(q)\cdot\boldsymbol{r}$, 
$b(q)=\sqrt{|\boldsymbol{d}(q)|^{2}-a(q)^2}$, and 
$\theta(q)=-\arctan\frac{\boldsymbol{d}(q)\cdot
\boldsymbol{n}}{\boldsymbol{d}(q)\cdot\boldsymbol{t}}$. 
Then the Hamiltonian in the LZ frame becomes 
\begin{align}
 \hat{\mathcal{H}}_{\mathrm{LZ}}(q)
   =&\hat{U}^{\dagger}\hat{\mathcal{H}}(q)\hat{U}
     -i\hat{U}^{\dagger}\partial_{t}\hat{U}\nonumber\\
   =&\Big[\Big(a(q)+\frac{\theta'(q)}{2}\frac{dq}{dt}\Big)
     \boldsymbol{r}+b(q)\boldsymbol{t}\Big]
     \cdot\hat{\boldsymbol{\sigma}}.
\label{eq:HamilLZframe_SM}
\end{align}
In the case of the model Eq.~\eqref{eq:Hamil_q2_SM}, 
the parameters are $a(q)=m$, $b(q)=vq$, 
and $\theta'(q)=-\kappa_{\parallel}v/2$. 
Through the transformation, 
the additional quadratic term is eliminated 
and the gap is effectively modified
from $m$ to $m_{\mathrm{eff}}=m+\kappa_{\parallel}vF/4$. 
The above formulation shows that
the geometric meaning of $\kappa_{\parallel}$ is
the curvature of $\boldsymbol{d}(q)$ in the plane spanned by 
$\boldsymbol{t}$ and $\boldsymbol{n}$ at $q=0$. 

With the application of the DDP 
method~\cite{Dykhne1962JETP,Davis1976JCP} 
for Eq.~\eqref{eq:HamilLZframe_SM}, 
the tunneling probability is expressed as
\begin{align}
 P\simeq\exp\Big[
   -2\mathrm{Im}\int_{0}^{q_{c}}
     \frac{\Delta(q)}{|F(q)|}dq\Big],
\label{eq:TunnelProbGeneral}
\end{align}
where 
$\Delta(q)=2[(a(q)-\theta'(q)F(q)/2)^{2}+b(q)^{2}]^{1/2}$ 
is the energy difference 
and $F(q)=-\frac{dq}{dt}$ is the Jacobian (expressed as function of $q$). 
In the DDP method, the integration path is deformed from the real axis, 
and the singular point closest to the real axis 
governs the tunneling probability. 
In Eq.~\eqref{eq:TunnelProbGeneral}, 
the integration is performed to $q_{\mathrm{c}}$ (on the imaginary axis), 
which is defined as a point in complex plane 
where the gap vanishes $\Delta(q_{\mathrm{c}})=0$ 
(the branching point of square root). 
For the linear sweep $q=-Ft$,
the Jacobian is just $F(q)=-\frac{dq}{dt}=F$.
Applying Eq.~\eqref{eq:TunnelProbGeneral} to 
the model Eq.~\eqref{eq:Hamil_q2_SM}, 
and noticing $a(q)=m$,
$b(q)=\sqrt{(vq)^{2}+(\kappa_{\parallel}v^{2}q^{2}/2)^{2}}=vq+\mathcal{O}(q^{3})$,
and
$\theta'(q)=-\frac{d}{dq}\arctan(\kappa_{\parallel}vq/2)
=-\kappa_{\parallel}v/2+\mathcal{O}(q^{2})$,
we can calculate the tunneling probability as 
\begin{align}
 P(F)=&\exp\bigg[-\frac{4}{|F|}
   \int_{0}^{\frac{1}{|v|}(m+\kappa_{\parallel}vF/4)}
     \sqrt{(m+\kappa_{\parallel}vF/4)^{2}-(vq)^{2}}dq\bigg]
\nonumber\\
   =&\exp\bigg[
      -\frac{\pi}{4}
        \frac{(2m+\kappa_{\parallel}vF/2)^{2}}{ v|F|}\bigg]
\label{eq:LZquadSM}
\end{align}
as given in Eq.~\eqref{eq:LZquad}.

\section{Twisted Schwinger effect in 2D: Nonadiabatic opto-valleytronics}
In the following sections, we study how nonadiabatic geometric effects 
in the tunneling probability Eq.~(\refeq{eq:LZquadSM2}) lead to nontrivial 
dynamics of electrons in Dirac and Weyl semimetals 
driven by strong electric laser fields.

We begin our analysis with the dynamics of 2D Dirac fermions in rotating electric fields. 
We introduce the field as gauge potential 
$\boldsymbol{A}=A(-\sin(\Omega t),\cos(\Omega t))$ 
[electric field $\boldsymbol{E}=E(\cos(\Omega t),\sin(\Omega t))$ 
($E=A\Omega>0$)], 
and the effective Hamiltonian for the 
fermions with chirality $\xi=\pm$ is given as 
\begin{align}
 \hat{\mathcal{H}}
   =v[\xi(k_{x}+eA_{x})\hat{\sigma}^{x}
     +(k_{y}+eA_{y})\hat{\sigma}^{y}]
     +m\hat{\sigma}^{z},
 \label{eq:laserHamiltonian}
\end{align}
where $e$ $(>0)$ is the elementary charge, 
$v$ is the Fermi velocity, and 
$m$ $(>0)$ is the mass parameter. 
This model has implication to valleytronics in 2D materials 
such as monolayer transition metal dichalcogenide (TMD) 
and graphene~\cite{Rycerz2007NatPhys,Schaibley2016NatRevMater},
where laser-induced valley polarization is 
demonstrated~\cite{Yao2008PRB,Xiao2012PRL,Mak2012NatNano,Cao2012NatComm,Zeng2012NatNano,Jones2013NatNano,KelardehPRB2016}. 
In these materials, the chirality $\xi$ corresponds to 
the valley index specifying the two Dirac points $K_{\xi}$ 
in the dispersion.

\begin{figure}[hbt]
\centering
\includegraphics[width=0.7\textwidth]{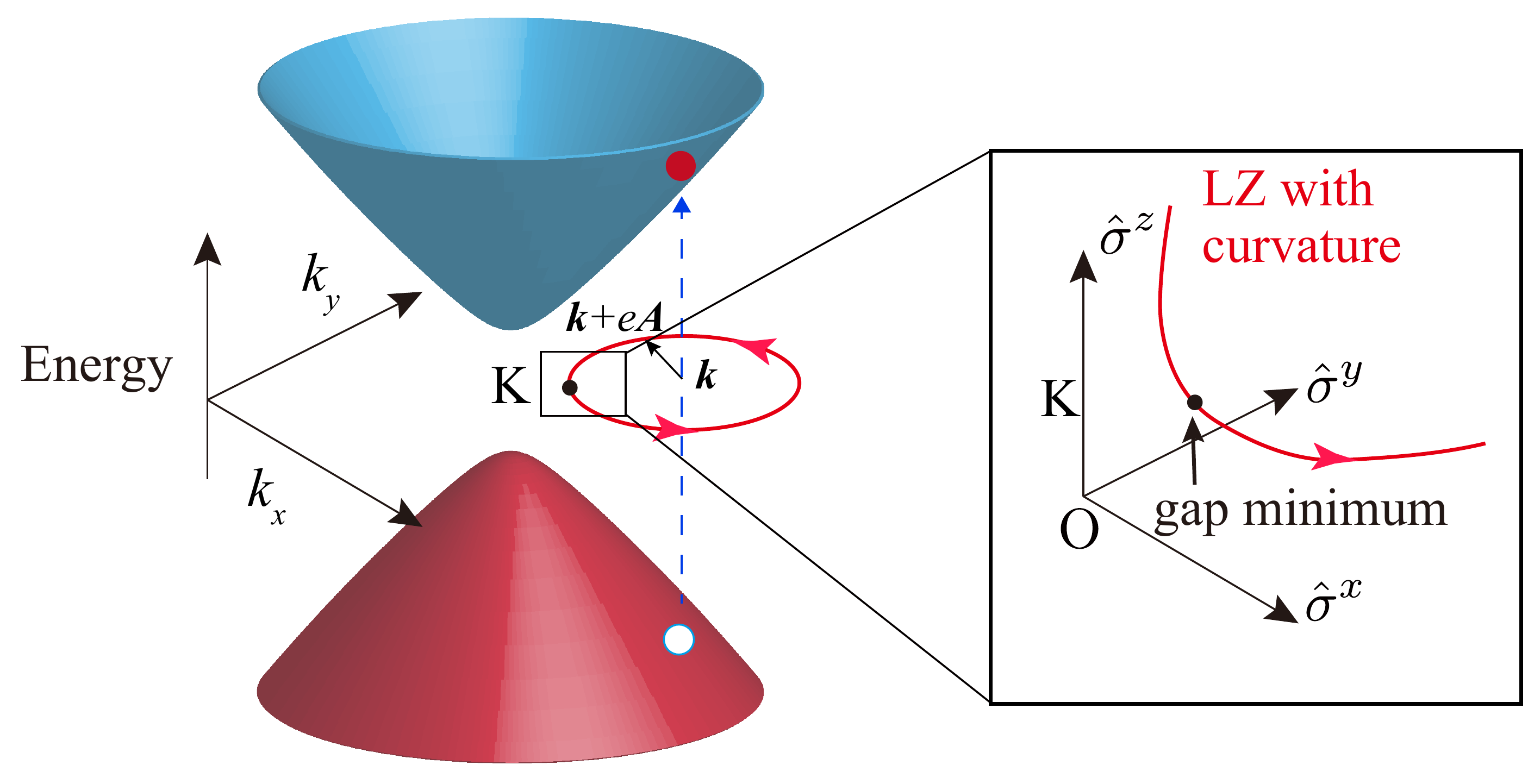}
\caption{
\textbf{Mapping from the twisted Schwinger effect to the twisted Landau Zener problem}:
In rotating electric fields, the electron-hole pairs have a covariant momentum 
$\boldsymbol{k}+e\boldsymbol{A}(t)$ which performs a rotating motion in the momentum space. 
During this dynamics, the energy gap minimizes 
when $\boldsymbol{k}+e\boldsymbol{A}(t)$ is closest to the $K$-point.  
We focus on this gap minimum point as depicted in the right box. 
By performing a quadratic expansion of the Hamiltonian $ \hat{\mathcal{H}}(t)$
in the time variable ($q=\Omega t$) around the gap minimum time, we 
obtain the twisted Landau Zener problem defined by Eq.~\eqref{eq:HamilGeneralOpSM}.
}
\label{fig:Circularmotion}
\end{figure}

We assume that the Fermi energy is zero,
and the time evolution starts from a zero-temperature ground state. 
After the field is switched on at $t=0$, 
nonadiabatic processes take place creating fermion-antifermion pairs. 
The tunneling process in momentum space can be 
mapped to a twisted Landau Zener problem discussed in the previous section 
as depicted in Fig.~\ref{fig:Circularmotion}.
We note that, in this mapping, we use a quadratic approximation. 
In this system, the laser frequency $\Omega$ plays 
the role of the speed parameter $F$ in the twisted LZ model. 
We allow $\Omega$ to be positive or negative 
which corresponds to the helicity specifying 
left or right circular polarization. 
The fermion-antifermion production probability per cycle of the laser field
is given by 
\begin{align}
\mathcal{P}_{\xi}(\boldsymbol{k})
   =\exp\Bigg[-\pi
   \frac{\displaystyle\Big(M-\frac{\xi\Omega m}{4M}\Big)^{2}}
     {veE}\Bigg],
\label{eq:P_MonoGra}
\end{align}
where we defined 
$M=\sqrt{v^{2}(|\boldsymbol{k}|-eE/(|\Omega|))^{2}+m^{2}}$. 
To derive this expression, we have 
expanded the Hamiltonian \eqref{eq:laserHamiltonian} 
around the time that minimizes the energy gap up to quadratic order 
obtaining the form
\eqref{eq:ABC} and used the tunneling formula Eq.~\eqref{eq:LZquad}.
We note that the remaining analysis is based on this approximate treatment (quadratic expansion) and 
the results are not exact. 

\begin{figure}[hbt]
\centering
\includegraphics[width=0.75\textwidth]{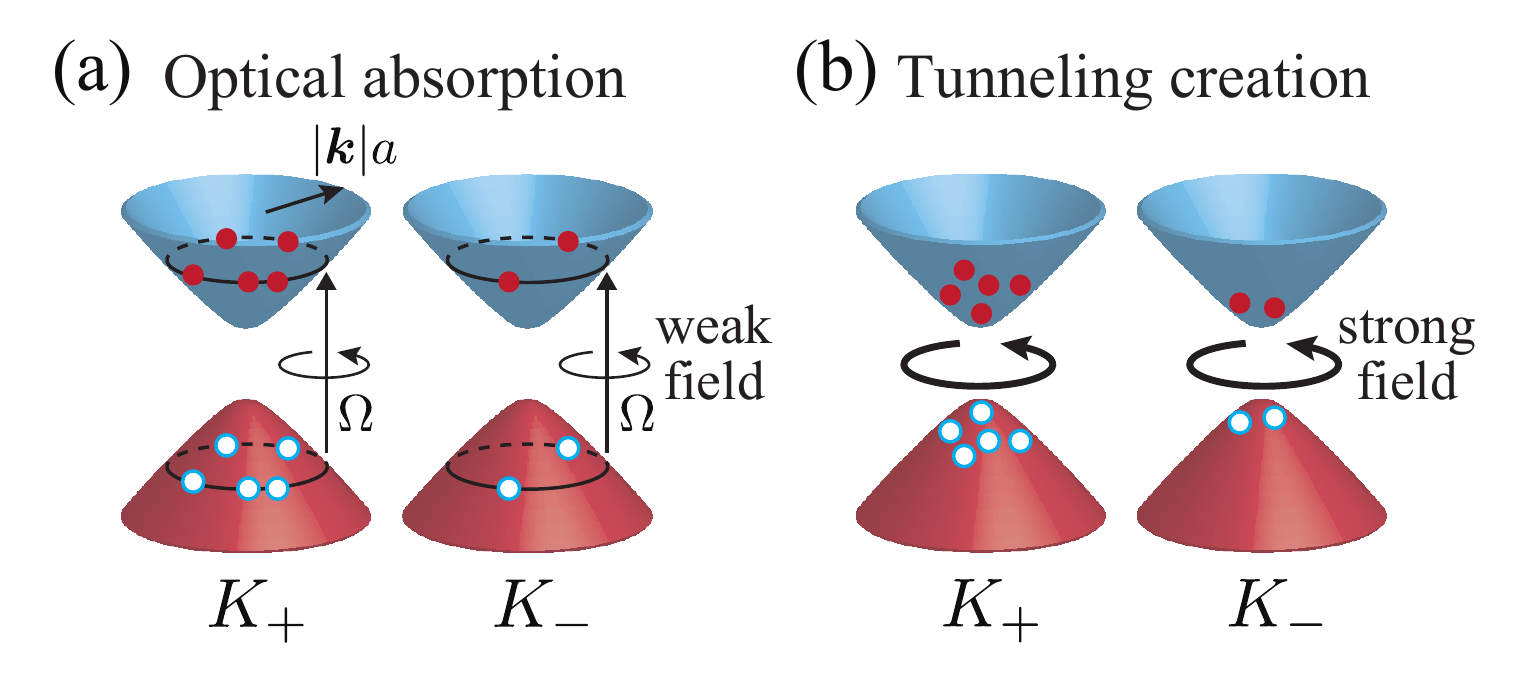}
\caption{
\textbf{Perturbative v.s. Nonadiabatic opto-valley polarization in 2D gapped Dirac fermion}:
Schematic picture of pair excitations 
 at the two valleys through two mechanisms. 
(a) In optical absorption, the pairs are 
concentrated on an equal energy curve $\Delta E(\boldsymbol{k})=\Omega$ 
due to energy conservation. 
 (b) In tunneling creation, 
the pairs are produced according to Eq.~\eqref{eq:P_MonoGra} (see Fig.~\ref{fig:CPL_kdepB}). 
}
\label{fig:CPL_kdepA}
\end{figure}

In Fig.~\ref{fig:CPL_kdepA}, we schematically compare 
the pair production induced by
(a) standard optical absorption process, and by (a) tunneling with nonadiabatic geometric effects. 
\begin{description}
\item[(a) Perturbative optical absorption process ]
In the case of standard optical absorption process, 
a perturbative picture of optical absorption is employed, where we consider the eigenstates $|\psi_n(\bs{k})\ket$
of the single body Hamiltonian $\hat{\mathcal{H}}(\bs{k})$ satisifying 
$\hat{\mathcal{H}}(\bs{k})|\psi_n(\bs{k})\ket
=E_n|\psi_n(\bs{k})\ket$. 
Electrons in the occupied bands are excited 
to the unoccupied bands, and the 
energy difference of the electron and hole is given by the photon energy. 
The momentum dependence of the excitation  density
is determined by the optical selection rule
encoded in the
transition dipole moment. 
In electrons in solids, 
the optical transition between 
bands $m$ and $n$ ($m\ne n$)
is given by the perturbation 
$\sum_{j=x,y,z} E^j(t)\mathcal{A}_{mn}^{j}(\bs{k})$
to the Hamiltonian $\hat{\mathcal{H}}(\bs{k})$.
The transition dipole moment 
is given by the Berry connection 
$\mathcal{A}_{mn}^{j}(\bs{k})=\bra\psi_m|i\pa_{j}|\psi_n\ket$
($j=x,y,z$)~\cite{SipePRB2000}. 
The photo absorption rate of circularly polarized laser in the $K_\xi$ valley
becomes 
$PA_{\xi=\pm}\propto |\mathcal{A}_{mn}^{\xi}(\bs{k})|^2|E(\Omega)|^2\delta(\Delta E-\Omega)$, 
where we defined $\mathcal{A}_{mn}^{\xi=\pm}(\bs{k})=\mathcal{A}_{mn}^{x}(\bs{k})\pm i\mathcal{A}_{mn}^{y}(\bs{k})$ and $\Delta E=E_c-E_v$~\cite{Yao2008PRB,Xiao2012PRL}. 

\item[(b) Nonadiabatic optical absorption process]
In the case of tunneling excitations, the properties of the 
excited pairs are different 
from the perturbative case. 
Energy is no longer conserved
since the Hamiltonian Eq.~\eqref{eq:laserHamiltonian} is depends on time.
Electron and hole pairs can be created even when their
energy difference is not equal to the photon energy. 
The role of the 
photo absorption rate $PA_{\xi=\pm}$
is now played by the production probability $\mathcal{P}_{\xi}(\boldsymbol{k})$ given in 
Eq.~\eqref{eq:P_MonoGra} or more generically in Eq.~\eqref{eq:LZquadSM2}.

\end{description}
 We summarize the comparison in Table~\ref{table:compare}.

\begin{table}[hbt]
\centering
\caption{{\bf Perturbative v.s. Nonadiabatic opto-valley polarization } }
\begin{tabular}{|c|c|}
\hline
Perturbative nonlinear optics~\cite{Yao2008PRB,Xiao2012PRL} &Nonadiabatic nonlinear optics (this work)\\
\hline
 Optical absorption  & Tunneling creation of electron-hole pairs \\
$PA_{\xi}\propto |\mathcal{A}_{mn}^{\xi}(\bs{k})|^2|E(\Omega)|^2\delta(\Delta E-\Omega)$&Tunneling probability $\mathcal{P}_{\xi}(\boldsymbol{k})
$ 
\\
Optical selection rule (transition dipole)& Geometric amplitude factor
\\
Energy momentum conservation modulo photon
&
Non-conservation of energy and momentum 
\\Valley polarization $\gamma=\frac{PA_{+}(\boldsymbol{k})}
         {PA_{-}(\boldsymbol{k})}$&
Valley polarization 
$\gamma=
\frac{\mathcal{P}_{+}(\boldsymbol{k})}
         {\mathcal{P}_{-}(\boldsymbol{k})}$
 \\
\hline
\end{tabular}
\label{table:compare}
\end{table}

\subsection{Valley polarization via tunneling creation}
\begin{figure}[htb]
\centering
\includegraphics[width=0.75\textwidth]{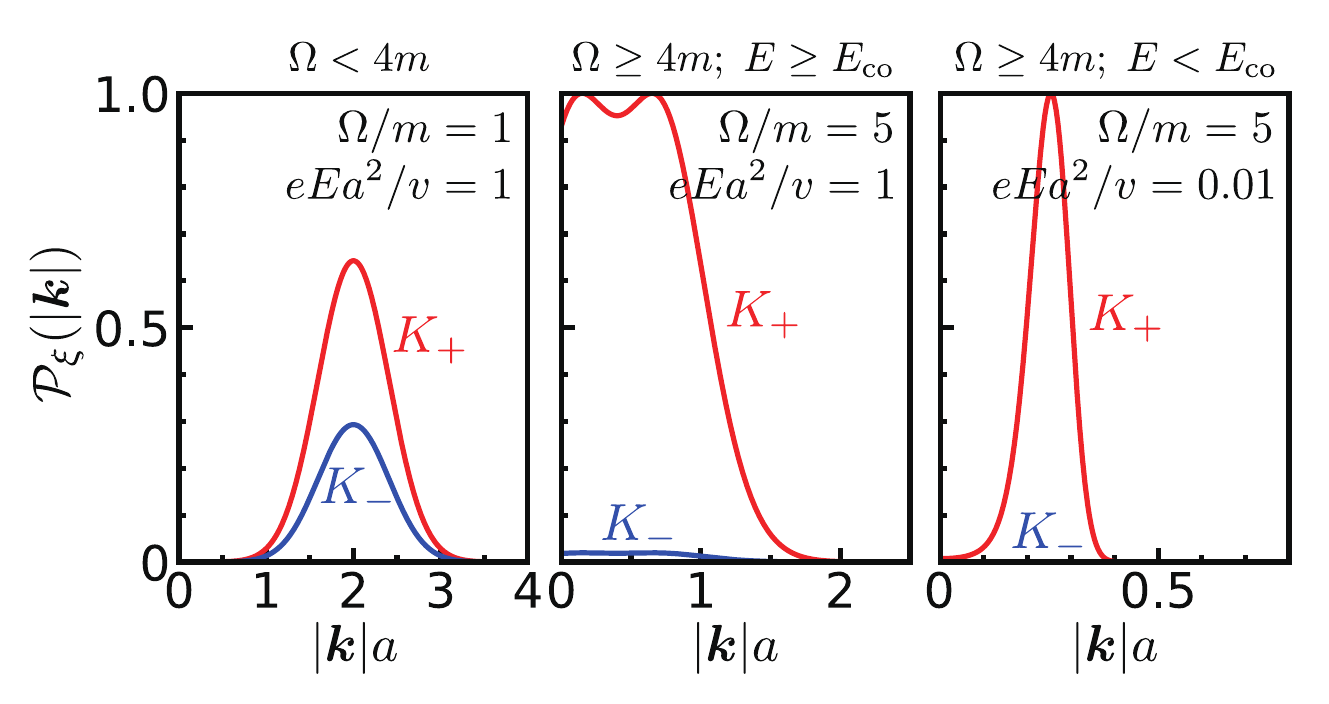}
\caption{
\textbf{Tunneling probability in 2D gapped Dirac fermion}:
The wavenumber dependence of 
the production probability 
$\mathcal{P}_{\xi}(\boldsymbol{k})$. 
The parameters are 
$(\Omega/m,eEa^{2}/v)=(1,1),(5,1),(5,0.01)$ 
and $ma/v=0.5$, 
where $a$ is the lattice constant. 
}
\label{fig:CPL_kdepB}
\end{figure}

In Fig.~\ref{fig:CPL_kdepB}, 
we plot the production probability for several $\Omega$. 
We see that there is a strong chirality dependence,
and the sign of $\xi\Omega$ determines whether 
excitations are ``optically allowed'' ($\xi\Omega>0$)
or ``optically forbidden'' ($\xi\Omega<0$). 
This difference originates from the geometric amplitude factor. 
In this sense, the optical selection rule\cite{Yao2008PRB,Xiao2012PRL} in 
perturbative optics is replaced by the nonadiabatic geometric effects
when nonperturbative strong field excitations are considered. 
The ratio of the production rates between the two chiralities 
\begin{align}
 \gamma
   =\frac{\mathcal{P}_{+}(\boldsymbol{k})}
         {\mathcal{P}_{-}(\boldsymbol{k})}
   =\exp\Big(\frac{\pi\Omega m}{veE}\Big)
\label{eq:PL}
\end{align}
is independent of the wavenumber. 
In the gapless case, as in graphene, $\gamma$ is unity 
and there is no valley dependence. 
When the gap parameter $m$ is finite, as in monolayer TMD, 
imbalance becomes finite and the ratio exponentially grows or decays 
with increasing $|\Omega|/E$.

Next, let us study the wavenumber dependence of the production probability
as depicted in Fig.~\ref{fig:CPL_kdepB}. 
The distribution is rotationally symmetric 
and only depends on $|\boldsymbol{k}|a$ ($a$: lattice constant). 
They have peaks 
as shown in Fig.~\ref{fig:CPL_kdepB}
at $|\boldsymbol{k}|=k_{\mathrm{peak}}$, where
\begin{empheq}[left={k_{\mathrm{peak}}=\empheqlbrace}]{align}
 & \frac{eE}{|\Omega|}
 && (\Omega<4m),
\label{eq:kpeak1}\\
 & \frac{eE}{|\Omega|}\pm
   \frac{1}{v}\sqrt{m(\xi\Omega/4-m)}
 && (\Omega\geq 4m;\; E\geq E_{\mathrm{co}}),
\label{eq:kpeak2}\\
 & \frac{eE}{|\Omega|}+
   \frac{1}{v}\sqrt{m(\xi\Omega/4-m)}
 && (\Omega\geq 4m;\; E< E_{\mathrm{co}}).
\label{eq:kpeak3}
\end{empheq}
The crossover field is defined by 
\begin{align}
E_{\mathrm{co}}=
   \frac{|\Omega|}{ev}\sqrt{m(\xi\Omega/4-m)}.
\label{eq:Eco}
\end{align}
 We can understand the peak structure 
from the wavenumber dependent effective mass parameter 
in Eq.~\eqref{eq:P_MonoGra} defined by 
\begin{align}
 m_{\mathrm{eff}}=
   M-\xi\Omega m/(4M). 
\label{eq:meff}
\end{align}
The peaks are dictated by the wavenumber 
minimizing the effective mass 
and their properties qualitatively change
depending on whether the frequency $\Omega$ 
is below or above $4m$. 
For $\Omega<4m$, the distributions have a single peak at the wavenumber
where $m_{\mathrm{eff}}>0$ is minimized.  
On the other hand, for higher frequencies $\Omega\geq 4m$, 
perfect tunneling 
takes place at the optically allowed valley ($\xi\Omega>0$)
when the effective gap $m_{\mathrm{eff}}$ close. 
There is a crossover when the electric field is increased,
The number of perfect tunneling peaks changes from 
one for $E<E_{\mathrm{co}}$ to two for $E\ge E_{\mathrm{co}}$. 
This field strength $E_{\mathrm{co}}$ characterizes a crossover 
of the total production rate which we will explain below.

\begin{figure}[t]
\centering
\includegraphics[width=0.8\textwidth]{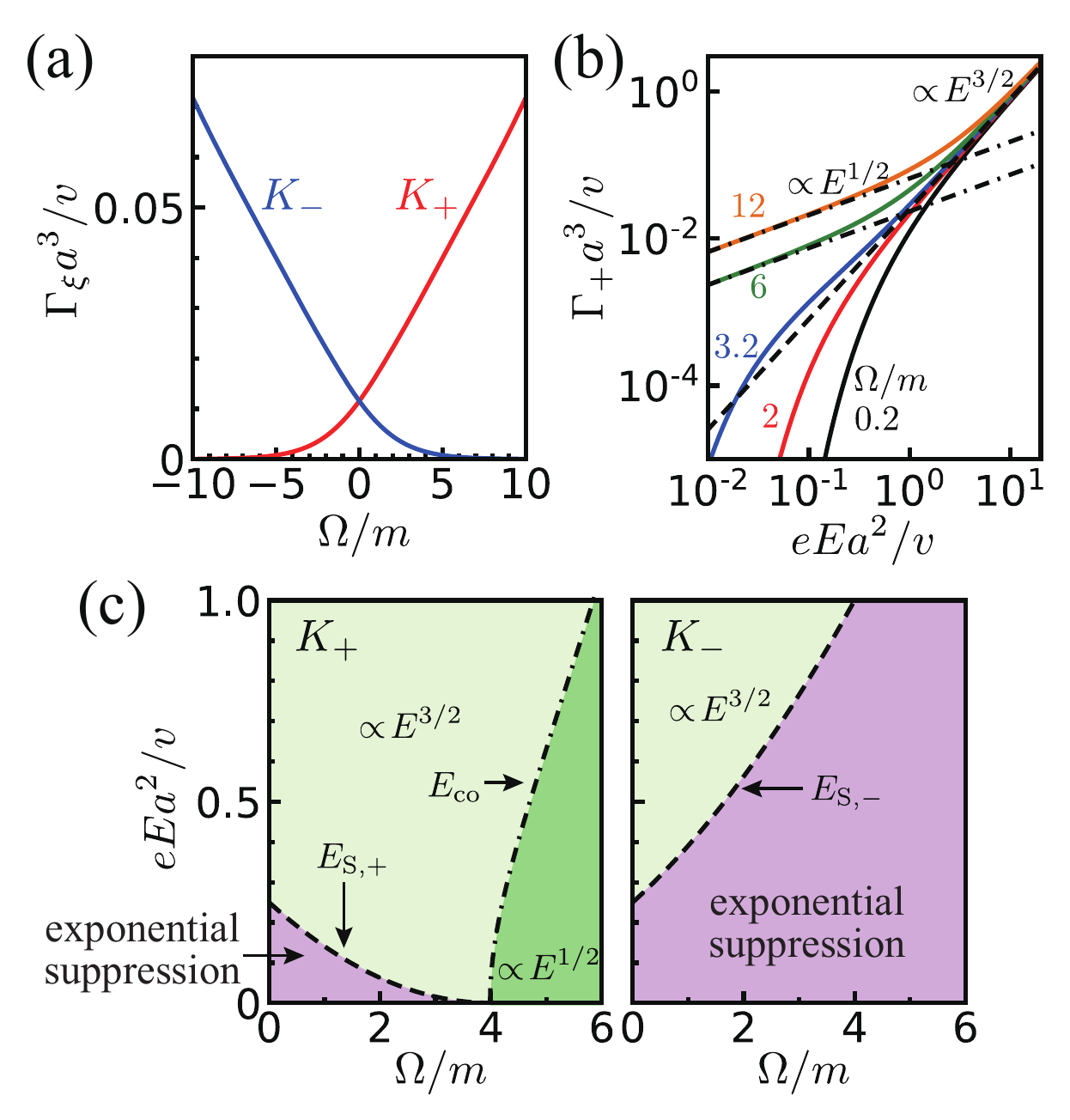}
\caption{
\textbf{2D gapped Dirac fermion}:
(a) The total pair production rate 
per unit of time and volume. 
We fix $ma/v=0.5$ and $eEa^{2}/v=1$. 
(b) The electric field dependence of 
the total production rate 
$\mathcal{P}_{\xi}^{\mathrm{tot}}$. 
(c) $(E,\Omega)$-phase diagram of 
the twisted Schwinger effect. 
}
\label{fig:CPL_Ptot}
\end{figure}

\subsection{Crossover in the production rate} 
We define the total fermion-antifermion production rate 
per unit of time and volume as 
$\Gamma_\xi\equiv \frac{|\Omega|}{(2\pi)^{3}}
\int d\boldsymbol{k} \mathcal{P}_{\xi}(\boldsymbol{k})$
and plot it against frequency in Fig.~\ref{fig:CPL_Ptot}(a). 
We see clearly the rectification effect where the imbalance ratio 
$\Gamma_{+}/\Gamma_{-}=\gamma$ 
increases exponentially for large $\Omega/E$ following Eq.~\eqref{eq:PL}. 
In the low-frequency region, 
it takes the form~(Appendix \ref{sec:2Ddetail})
\begin{align}
 \Gamma_{\xi}\simeq&
   \frac{eE}{(2\pi)^{2}}\sqrt{\frac{eE}{v}}
     \exp\Big(-\pi\frac{E_{\mathrm{S},\xi}}{E}\Big).
\label{eq:Rate_Schwin} 
\end{align}
Here we define the Schwinger limit of field strength as 
\begin{align}
 E_{\mathrm{S},\xi}\equiv
   (\overline{m}_{\mathrm{eff},\xi})^{2}/(ve)=
     (m-\xi \Omega /4)^2/(ve),
\label{eq:Schwingerlimit}
\end{align}
where $\overline{m}_{\mathrm{eff},\xi}$
is the effective mass at the gap minimizing wavenumber 
Eq.~\eqref{eq:kpeak1}. 
Equation~\eqref{eq:Rate_Schwin} is an extension 
of Schwinger's production rate 
evaluated originally for a DC electric field 
to the case of rotating electric field. 
For $\Omega=0$, Eq.~\eqref{eq:Rate_Schwin} coincides with the 2D version 
of Schwinger's result~\cite{Schwinger1951PR,OkaAoki2005PRL}
with the QED Schwinger limit $E_{\mathrm{S}}=m_{e}^{2}c^{3}/(\hbar e)$ 
obtained by replacing $m\to m_{e}c^{2}$ and $v\to \hbar c$. 

Figure \ref{fig:CPL_Ptot}(b) shows the electric field dependence of the 
production rate with the optically allowed chirality ($\xi\Omega>0$) 
for several frequencies. 
For strong fields, all curves converge to the dashed line 
$\Gamma_{\xi}\to\frac{eE}{(2\pi)^{2}}\sqrt{\frac{eE}{v}}$
described by the asymptotic form of 
Eq.~\eqref{eq:Rate_Schwin} independent of $\Omega$. 
For weak fields, we observe two different behaviors. 
The low frequency ($\Omega <4m$) curves 
drop below the dashed line following 
Eq.~\eqref{eq:Rate_Schwin} 
due to the exponential suppression of tunneling at weak fields.  
In contrast, curves for high frequency ($ \Omega \geq 4m$)
turn above the dashed line 
and converge to a $\Gamma_{+}\propto E^{1/2}$ behavior. 
In Fig.~\ref{fig:CPL_Ptot}(c), 
we summarize the tunneling behaviors into 
a $(E,\Omega)$-phase diagram, which we explain below. 


\paragraph{Low frequency ($|\Omega|<4m$) (Appendix \ref{sec:2Ddetail1})}
The Schwinger limit $E=E_{\mathrm{S},\xi}$ 
[Eq.~\eqref{eq:Schwingerlimit}]
characterizes the crossover from the weak field 
exponentially suppressed regime to the 
$\Gamma_{\xi}\propto E^{3/2}$ behavior at strong field. 
Increasing $\Omega$ from zero, 
the Schwinger limit $E_{\mathrm{S},\xi}$ for the optically allowed chirality ($\xi\Omega>0$)
decreases and becomes zero 
at $|\Omega|=4m$, where perfect tunneling starts to happen. 
In contrast, for the optically forbidden chirality ($\xi\Omega<0$),
 $E_{\mathrm{S},\xi}$ monotonically increase 
against $|\Omega|$. 
This suppression of tunneling is the consequence of 
counterdiabaticity in the twisted LZ tunneling. 

\paragraph{High frequency ($|\Omega| \geq 4m$) (Appendix \ref{sec:2Ddetail2})} 
For the optically allowed chirality $\xi\Omega>0$, 
the effective gap closes 
and the Schwinger limit vanishes due to perfect tunneling. 
There is a crossover taking place 
around $E=E_{\mathrm{co}}$ defined in Eq.~\eqref{eq:Eco} where the number of the peaks in the distribution 
function changes (Fig.~\ref{fig:CPL_kdepB}).  The production rate shows 
the $\Gamma_{\xi}\propto E^{3/2}$ behavior at strong field $E>E_{\mathrm{co}}$ 
and changes to 
a $\Gamma_{\xi}\propto E^{1/2}$ behavior at weak fields $E<E_{\mathrm{co}}$ . 
In particular, in the weak field regime, the production rate
shows an asymptotic form
\begin{align}
 \Gamma_{\xi}\simeq&
   \frac{|\Omega|}{(4\pi)^{2}v}
     \sqrt{|\Omega|m}
     \sqrt{\frac{eE}{v}}
\label{eq:Rate_PT}
\end{align}
 for optically allowed $\xi$,
which is evaluated in the appendix\ref{sec:2Ddetail2}. 
On the other hand, for the optically forbidden chirality $\xi\Omega<0$,
the Schwinger limit monotonically increases as $|\Omega|$ increases. 

Before closing this section, 
we give an estimate of the Schwinger limit in solid-state materials. 
The Fermi velocity $v$ and mass parameters $m=\Delta/2$
for a typical TMD material MoS$_2$ is given by 
$v=3.5\;\AA \mbox{eV}$ and $m=\frac{\Delta}{2}=0.83\;\mbox{eV}$ where $\Delta$ is the optical gap~\cite{Xiao2012PRL}. 
The Schwinger limit of MoS$_2$  is given by 
$E_{S,\xi}(0)=m^2/(ve)=0.20\;\mbox{V}/\AA=2.0\times 10^9\;\mbox{V/m}$
for $\Omega=0$. 
For finite photon energy $\Omega$, the Schwinger limit decreases and vanish at 
$\Omega=4m = 2\Delta=3.3\;\mbox{eV}$ for one valley. 
Experimentally realizable fields using
THz laser ($\Omega\sim 0$) is around $E_{\rm THz}=10^8 \;\mbox{V/m}$~\cite{HiroriAPL2011}, 
while it exceeds $E_{\rm NI}=10^9\; \mbox{V/m}$~\cite{KawakamiNatPhoto2018} in the near-infrared region $\Omega=0.6-1\;\mbox{eV}$.
Thus, the field strength of near-infrared lasers is comparable to 
the Schwinger limit at finite photon energy $\Omega$ and can be used to verify our predictions. 


\section{Twisted Schwinger effect in 3D: Nonadiabatic photo-current}

Next, we proceed to an analysis of 3D massless Dirac fermions 
subject to rotating electric fields described by the Hamiltonian 
\begin{align}
 \hat{\mathcal{H}}_{\mathrm{3D}}
   =&v\sum_{j=x,y,z}\hat{\gamma}^{0}\hat{\gamma}^{j}(q_{j}+eA_{j})
\nonumber\\
   =&
\begin{pmatrix}
 -v\sum_{j=x,y,z}(q_{j}+eA_{j})\hat{\sigma}^{j}
 & 0 \\
 0 &
 v\sum_{j=x,y,z}(q_{j}+eA_{j})\hat{\sigma}^{j}
\end{pmatrix}
\label{eq:laserHamiltonian3D}
\end{align}
with the 3D wave number 
$\boldsymbol{q}=(\boldsymbol{k},k_{z})$, 
$\boldsymbol{A}=A(-\sin(\Omega t),\cos(\Omega t),0)$ 
and the gamma matrices 
\begin{align}
 \hat{\gamma}^{0}=
\begin{pmatrix}
 0 & I \\
 I & 0
\end{pmatrix}
,\quad
\hat{\gamma}^{j}=
\begin{pmatrix}
 0 & \hat{\sigma}^{j} \\
 -\hat{\sigma}^{j} & 0
\end{pmatrix}
\;(j=x,y,z).
\nonumber
\end{align}

We can recast this Hamiltonian to the 2D Dirac Hamiltonian studied in the previous section. 
By performing the unitary transform 
\begin{align}
 \hat{U}=
\begin{pmatrix}
 \exp(i\frac{\pi}{2}\hat{\sigma}^{x}) & 0 \\
 0 & I
\end{pmatrix}
\nonumber
\end{align}
to the Hamiltonian Eq.~\eqref{eq:laserHamiltonian3D}, 
we obtain 
\begin{align}
 \hat{U}^{\dagger}\hat{\mathcal{H}}_{\mathrm{3D}}\hat{U}=
\begin{pmatrix}
 \hat{\mathcal{H}}_{-} & 0 \\
 0 & \hat{\mathcal{H}}_{+}
\end{pmatrix}
,\nonumber
\end{align}
where 
\begin{align}
 \hat{\mathcal{H}}_{\xi}=v
   [\xi(k_{x}-eA\sin(\Omega t))\hat{\sigma}^{x}
      +(k_{y}+eA\cos(\Omega t))\hat{\sigma}^{y}
      + k_{z}\hat{\sigma}^{z}]
\end{align}
is the Weyl Hamiltonian with chirality $\xi=\pm$. 
This Hamiltonian is equivalent to the 
2D Dirac Hamiltonian Eq.~\eqref{eq:laserHamiltonian}
studied in the previous section with the replacement 
of the mass $m$ by $vk_{z}$. 
Thus, 
the fermion-antifermion production probability per cycle of the laser field
is given by 
\begin{align}
\mathcal{P}_{\xi}(\boldsymbol{k})
   =\exp\Bigg[-\pi
   \frac{\displaystyle\Big(M-\frac{\xi\Omega vk_{z}}{4M}\Big)^{2}}
     {veE}\Bigg],
\label{eq:P_3D}
\end{align}
where we defined 
$M=v\sqrt{(|\boldsymbol{k}|-eE/|\Omega|)^{2}+k_{z}^{2}}$.

Below, we assume that the Fermi energy is at the Dirac point
and exploit the scaling symmetry 
rewriting the model with variables 
$\tilde{t}=|\Omega|t$ and 
$\tilde{\boldsymbol{q}}=v\boldsymbol{q}/|\Omega|$.
The Schr\"odinger equation is recast to 
 $i\partial_{\tilde{t}}
   |\Psi_{\xi}(\tilde{t})\rangle
   =\hat{\tilde{\mathcal{H}}}_{\xi}
     |\Psi_{\xi}(\tilde{t})\rangle$
with 
 $\hat{\tilde{\mathcal{H}}}_{\xi}
   =\xi(\tilde{k}_{x}-\mathrm{sgn}(\Omega)\tilde{A}\sin\tilde{t})
       \hat{\sigma}^{x}
      +(\tilde{k}_{y}+\tilde{A}\cos\tilde{t})\hat{\sigma}^{y}
      + \tilde{k}_{z}\hat{\sigma}^{z}.$
Then we can set the frequency $|\Omega|$ to unity and 
\begin{align}
\tilde{A}=veA/|\Omega|=veE/\Omega^{2}
\label{eq:Atilde}
\end{align}
is the unique scaling parameter that
characterizes the field strength.

\begin{figure*}[tbh]
\centering
\includegraphics[width=0.94\textwidth]{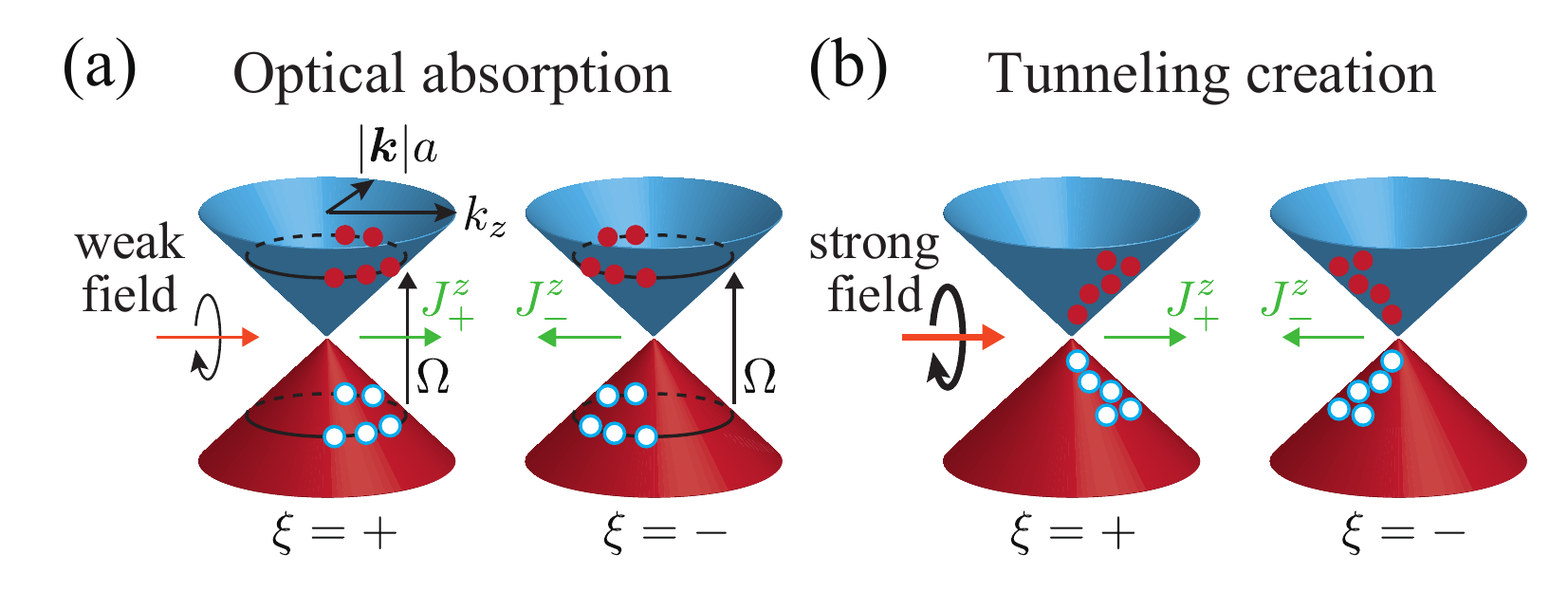}
\caption{
\textbf{Nonadiabatic v.s. Perturbative photo-current generation in 3D massless Dirac fermion}: 
Schematics of pair production and resulting photocurrent 
in the two Weyl components induced by the 
(a) tunneling creation and (b) optical excitations. 
 (a) In tunneling creation, the pairs are produced according to Eq.~\eqref{eq:P_3D}. See Fig.~\ref{fig:Dirac3D} for the results. 
(b) In optical absorption, the pairs are concentrated on equal energy curves due to energy conservation. 
}
\label{fig:Schematic3D}
\end{figure*}

In Fig.~\ref{fig:Schematic3D}, 
we schematically compare 
the pair production in the 3D Dirac systems 
induced by 
 (a) standard optical absorption process to (b) tunneling creation. 
Similarly to the 2D case summarized in Fig.~\ref{fig:CPL_kdepA} 
and Table~\ref{table:compare},
the nonadiabatic geometric effects take the place of 
the optical selection rule~\cite{Chan2017PRB}. 
Circularly polarized laser field propagating along the $z$ axis 
induces vertical transitions
that are imbalanced between $\pm k_{z}$. The imbalance in the created 
pairs result in a photocurrent $J_{\xi}^{z}$ 
for each Weyl component $\xi=\pm$ as we will see below.
In Dirac systems with chiral and mirror reflection symmetries, 
the total photocurrent cancels since 
the production rate 
$\mathcal{P}_{\xi}(\tilde{\boldsymbol{q}})$ is symmetric under $\xi\to -\xi$, $k_{z}\to -k_{z}$.
On the other hand, if these symmetries 
are broken, it is possible to realize finite $U(1)$ photocurrent in 
a similar way as in the optical absorption mechanism 
proposed in \cite{Chan2017PRB,Ma2017NatPhys,Chang2020PRL}. 

\subsection{Expression of the total and chiral current}
The total U(1) and chiral current operators 
are represented as 
\begin{align}
 \hat{J}^{z}=-ve\hat{U}^{\dagger}\hat{\gamma}^{0}
   \hat{\gamma}^{z}\hat{U}=-ve
\begin{pmatrix}
 \hat{\sigma}^{z} & 0 \\
 0 & \hat{\sigma}^{z}
\end{pmatrix}
,\nonumber
\end{align}
and 
\begin{align}
 \hat{J}^z_5=-ve\hat{U}^{\dagger}\hat{\gamma}_{5}
   \hat{\gamma}^{0}\hat{\gamma}^{z}\hat{U}=-ve
\begin{pmatrix}
 -\hat{\sigma}^{z} & 0 \\
 0 & \hat{\sigma}^{z}
\end{pmatrix}
,\nonumber
\end{align}
where $\hat{\gamma}_{5}=
\begin{pmatrix}
-I & 0 \\
 0 & I
\end{pmatrix}$. 
To evaluate the expectation value of the currents, 
we need to estimate the distribution of the electron-hole pairs. 
This can be done by calculating the time evolution of the density matrix 
in the presence of relaxation. 
One of the schemes is to employ the 
Liouville von Neumann equation in the momentum space 
within the relaxation time approximation~\cite{Sato2019PRB}. 
However, for simplicity, here we assume that the system is on-shell, 
i.e., the  density matrix is diagonal in the 
eigenstate basis and the distribution is obtained by the balance between the 
creation process characterized by the 
tunneling probability and the relaxation time. 
This can be done by first representing the 
 density matrix as
$\rho_{\boldsymbol{q},\xi}(t)
=n_{\boldsymbol{q},\xi}(t)|\Psi_{\boldsymbol{q},\xi,1}\rangle
\langle\Psi_{\boldsymbol{q},\xi,1}|
+[1-n_{\boldsymbol{q},\xi}(t)]
|\Psi_{\boldsymbol{q},\xi,2}\rangle\langle\Psi_{\boldsymbol{q},\xi,2}|$, 
where $|\Psi_{\boldsymbol{q},\xi,1}\rangle$ and 
$|\Psi_{\boldsymbol{q},\xi,2}\rangle$ are 
the states for upper and lower bands with chirality $\xi$. 
The master equation is 
\begin{align}
\frac{dn_{\boldsymbol{q},\xi}(t)}{dt}=[1-n_{\boldsymbol{q},\xi}(t)]
\mathcal{P}_{\xi}(\boldsymbol{q})\frac{|\Omega|}{2\pi}
-n_{\boldsymbol{q},\xi}(t)/\tau,
\end{align}
where $\tau$ is the relaxation time. Note that we have the factor $\frac{|\Omega|}{2\pi}$ (= inverse of the time period) in the first term on the r.h.s. since $\mathcal{P}_{\xi}(\boldsymbol{q})$
is defined as the tunneling probability per cycle. 
In the steady state $dn_{\boldsymbol{q},\xi}(t)/dt=0$, 
if we assume that the relaxation time is short 
$|\Omega|\tau\mathcal{P}_{\xi}(\boldsymbol{q})/(2\pi)\ll 1$, 
we obtain 
$n_{\boldsymbol{q},\xi}(t)
=|\Omega|\tau\mathcal{P}_{\xi}(\boldsymbol{q})/(2\pi)$. 
The current density for the component of chirality $\xi$ 
is provided as 
\begin{align}
 J_{\xi}^{z}=-2ve\frac{|\Omega|\tau}{2\pi a^{3}}
   \Big(\frac{a}{2\pi}\Big)^{3}
     \int d\boldsymbol{q}
     \frac{k_{z}}{\sqrt{|\boldsymbol{k}|^{2}+k_{z}^{2}}}
     \mathcal{P}_{\xi}(\boldsymbol{q})
 =-\frac{2e\tau|\Omega|^{4}}{(2\pi)^{4}v^{2}}
   \int d\tilde{\boldsymbol{q}}
   \frac{\tilde{k}_{z}}
     {\sqrt{|\tilde{\boldsymbol{k}}|^{2}+\tilde{k}_{z}^{2}}}
     \mathcal{P}_{\xi}(\tilde{\boldsymbol{q}}).
\end{align}
We can calculate the total and chiral (spin) currents as 
$J^{z}=J_{+}^{z}+J_{-}^{z}$ and 
$J_{5}^{z}=J_{+}^{z}-J_{-}^{z}$.

We also define total (chiral) production rates as 
$\Gamma_{\mathrm{tot}}^{\mathrm{3D}}=
\Gamma_{+}^{\mathrm{3D}}+\Gamma_{-}^{\mathrm{3D}}$ 
($\Gamma_{5}^{\mathrm{3D}}=\Gamma_{+}^{\mathrm{3D}}-\Gamma_{-}^{\mathrm{3D}}$)
using
$\Gamma_{\xi}^{\mathrm{3D}}=\frac{|\Omega|^{4}}{(2\pi)^{4}v^{3}}
\int d\tilde{\boldsymbol{q}}\mathcal{P}_{\xi}(\tilde{\boldsymbol{q}})$.
Due to the symmetry of $\mathcal{P}_{\xi}(\tilde{\boldsymbol{q}})$ under $\xi\to -\xi$, $k_{z}\to -k_{z}$, 
$\Gamma_{5}^{\mathrm{3D}}=J^{z}=0$ holds. 

\subsection{Novel crossover between weak-to-strong field behaviors} 
Now, let us discuss the physical consequence of the geometric nonadiabatic 
effect in the tunneling creation in 3D Dirac fermions. 
In the massless Dirac and Weyl fermions, 
there is no tunneling threshold and we expect that the 
 total production rate shows a power-law behavior
against the electric field strength. 
We show that there is a crossover 
between the weak and strong field regimes
accompanied by a change in power. 

\begin{figure*}[t]
\centering
\includegraphics[width=0.98\textwidth]{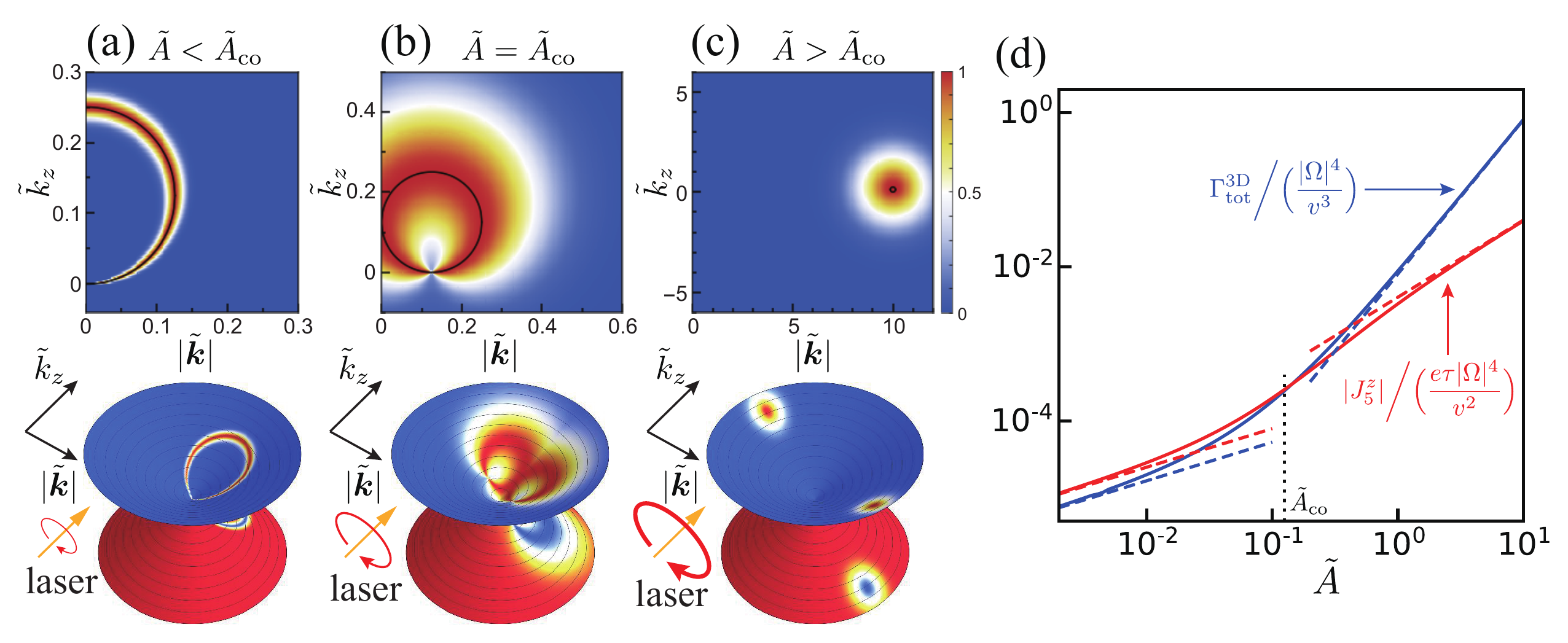}
\caption{
\textbf{3D massless Dirac fermion}:
(a)-(c) 
The production probability $\mathcal{P}_{\xi}(\tilde{\boldsymbol{q}})$ 
for chirality $\xi=+$ plotted
for several field strength parameters 
(a) $\tilde{A}=veE/\Omega^{2}=0.001$,
(b) $\tilde{A}=1/8$, and 
(c) $\tilde{A}=10$. 
They are rotationally symmetric around the $\tilde{k}_{z}$ axis
and the probability for particles with chirality $\xi=-$ is reflected 
as $\tilde{k}_{z}\to -\tilde{k}_{z}$. 
The solid black curve denotes wavenumber
at which perfect tunneling occurs 
[Eqs.~\eqref{eq:kpeak2} and \eqref{eq:kpeak3} with $m$ replace by $k_{z}$]. 
The lower panels show the fermion-antifermion pairs
on the Weyl cone $E=\pm\sqrt{|\tilde{\boldsymbol{q}}|}$ 
for fixed  $\tilde{k}_{y}=0$.
(d) The total production rate and chiral current are plotted
as blue and red solid curves while the dashed lines represent 
their asymptotic power-law behavior 
Eqs.~\eqref{eq:Gammacrossover} and \eqref{eq:Jcrossover}. 
}
\label{fig:Dirac3D}
\end{figure*}

In Figs.~\ref{fig:Dirac3D}(a)-\ref{fig:Dirac3D}(c), 
we plot the production probability 
$\mathcal{P}_{\xi}(\tilde{\boldsymbol{q}})$ 
for $\xi=+$ and $\Omega>0$ obtained in Eq.~\eqref{eq:P_3D}. 
It is rotationally symmetric around the $\tilde{k}_{z}$ axis. 
The production probability for the other chirality $\xi=-$ 
is a reflection of $\xi=+$ around the $\tilde{k}_{z}=0$ plane. 
The production probability shows peaks around the wavenumber 
satisfying the perfect tunneling conditions 
Eqs.~\eqref{eq:kpeak2} and \eqref{eq:kpeak3}
with $m$ replace by $k_z$.  
In the plane of $(|\tilde{\boldsymbol{k}}|,\tilde{k}_{z})$, 
the perfect tunneling peaks 
define a circle centered at 
$(|\tilde{\boldsymbol{k}}|,\tilde{k}_{z})=(\tilde{A},1/8)$ 
with a radius $1/8$ and are
 plotted as black solid curves. 
We find a crossover in the shape of the perfect tunneling peaks that 
occurs at 
\begin{align}
\tilde{A}_{\mathrm{co}}=1/8.
 \label{eq:Acrossover}
\end{align}
For $\tilde{A}<\tilde{A}_{\mathrm{co}}$ the circle is incomplete and 
approaches a semicircle in the small $\tilde{A}$ limit, and for 
large field $\tilde{A}\geq \tilde{A}_{\mathrm{co}}$ the circle becomes complete. 
Remembering the definition of $\tilde{A}$ given in Eq.~\eqref{eq:Atilde},
the crossover field strength is 
$E_{\mathrm{co}}=\frac{1}{8}\frac{(\hbar \Omega)^2}{e\hbar v}$,
where we have temporally recovered the Planck constant. 
For example, in the case of Cd$_{3}$As$_{2}$, the velocity parameter 
is of the order of $v\sim 10^5\mbox{m/s}$ and using $\hbar=6.6\times 10^{-16}\mbox{eVs}$, the 
crossover fields for photon energies $\hbar \Omega=1\mbox{eV}$ 
and $\hbar \Omega=1\mbox{meV}$
are $E_{\mathrm{co}}\sim 2\times 10^9\mbox{V/m}$ and 
 $E_{\mathrm{co}}\sim 2\times 10^3\mbox{V/m}$, respectively. 
 We stress that these parameters for the laser strength are experimentally feasible.

Next, we investigate how this crossover is seen in the physically observable quantities. 
We plot the total production rate $\Gamma_{\xi}^{\mathrm{3D}}$ 
and the chiral current in the $z$ direction $J_{5}^{z}$ in Fig.~\ref{fig:Dirac3D}(d). 
The quantities show a power-law behavior 
in the weak and strong field limits with different powers. 
The change of the power occurs 
around the crossover field $\tilde{A}=\tilde{A}_{\mathrm{co}}$ 
and their asymptotic behaviors are given by 
\begin{align}
 &\Gamma_{\mathrm{tot}}^{\mathrm{3D}}\bigg/
   \Big(\frac{|\Omega|^{4}}{v^{3}}\Big)\to
\begin{cases}
 \frac{1}{3(4\pi)^3}\tilde{A}^{1/2} &
 (\tilde{A}/\tilde{A}_{\mathrm{co}}\ll 1)\\
 \frac{2}{(2\pi)^3}\tilde{A}^{2} &
 (\tilde{A}/\tilde{A}_{\mathrm{co}}\gg 1),
\end{cases}
\label{eq:Gammacrossover}
\\
 &J_{5}^{z}\bigg/
   \Big(\frac{-e\tau|\Omega|^{4}}{v^{2}}\Big)\to
\begin{cases}
 \frac{\mathrm{sgn}(\Omega)}{2(4\pi)^3}\tilde{A}^{1/2} &
 (\tilde{A}/\tilde{A}_{\mathrm{co}}\ll 1) \\
 \frac{\mathrm{sgn}(\Omega)}{(2\pi)^3}\tilde{A}^{1} &
 (\tilde{A}/\tilde{A}_{\mathrm{co}}\gg 1).
\end{cases}
\label{eq:Jcrossover}
\end{align}
It is possible to analytically evaluate the asymptotic behaviors 
using the fact that the 
distribution around the peak is a Gaussian 
with a width scaling as $\tilde{A}^{1/2}$. 
The detailed calculation is given in appendix~\ref{sec:appendix3D}.

This is a novel nonperturbative crossover that originated from the 
the nonadiabatic geometric effect that has no 
perturbative analogue. 
Let us discuss how we can measure the current as well as the crossover 
in solid-state experiments.  
We have discussed a general theory based on Weyl and Dirac Hamiltonians. 
There are various material realizations of Weyl and Dirac Hamiltonians~\cite{RevModPhys.90.015001}, 
where the chirality $\xi$ may correspond to degrees of freedom such as
orbitals and spins as well as their mixtures. 
For example, in Co$_3$Sn$_2$S$_2$ \cite{Liu2018NatPhys}
the chirality $\xi$ corresponds to spin \cite{Ozawa2019JPSJ}
and the chiral current $J_{5}^{z}$ can be detected as a spin current.
The generation of $U(1)$ photocurrent due to optical absorption 
in Weyl semimetals with broken symmetry have been 
studied in refs.~ \cite{JuanNatCom2017,Chan2017PRB,Ma2017NatPhys,Chang2020PRL,OsterhoudtNatMat2019}. 
If the Fermi energy is non-zero and the system has finite 
carrier density,  the nonlinear anomalous Hall current~\cite{Sodemann2015PRL} can also contribute 
to the current generation in the $z$-direction~\cite{dantas2020nonperturbative}. 
The three mechanisms, i.e., tunneling creation, optical absorption, nonperturbative Hall current,
have different dependencies on the laser and material parameters such as Field strength, photon energy, and Fermi energy. 
The asymptotic behaviors of the physical observables in Eq.~\eqref{eq:Gammacrossover},\;\eqref{eq:Jcrossover} is useful in identifying the origin of the photoinduced current.


\section{Conclusion}
We studied the nonadiabatic geometric effects 
in quantum tunneling and found that they provoke
anomalous phenomena such as rectification, perfect tunneling
and counterdiabaticity. 
We derived the tunneling formula describing these effects through the 
modulation of the effective mass. 
We studied the implication of nonadiabatic geometric effects 
in the Schwinger effect, i.e., tunneling creation of carriers, induced by rotating electric fields. 
Two condensed matter applications are 
mentioned. One is the valley polarization that can be induced in 2D Dirac materials,
and the other is the generation of spin (and charge) current in 
3D Dirac (and Weyl) materials. 
Our finding adds another example to the 
rich nonperturbative phenomena
induced by circularly polarized laser in electronic systems~\cite{Oka2009PRB,Kitagawa2011PRB,Lindner2011NatPhys,
Wang2014EPL,Ebihara2016PRB,Chan2016PRL,Bucciantini2017PRB}. 
Finally, we comment that the interplay between the nonadiabatic geometric effects
and interaction 
is an open problem calling for further study.  
We point out that there is an interesting resemblance 
between the phase diagram of the twisted Schwinger effect
[Fig.~\ref{fig:CPL_Ptot}(c)] and that of a strongly interacting holographic 
model~\cite{Hashimoto2017JHEP,Kinoshita2018JHEP}.

\section*{Acknowledgements}

We would like to thank Masamitsu Hayashi, Ryo Shimano, Sota Kitamura, Takahiro Morimoto, 
Masafumi Udagawa, Francesco Peronaci, Alexandra Landsman, 
Hamed Koochaki Kelardeh, and Lisa Ortmann for fruitful discussions. 


\paragraph{Funding information}

This work was supported by JSPS KAKENHI
Grant No. JP21K03412 
and JST CREST Grant No. JPMJCR19T3, Japan.
J. W. acknowledges additional support from a Shanghai talent program. 
The work at Shanghai Jiao Tong
University is sponsored by Natural Science Foundation of Shanghai with Grant No. 20ZR1428400 and Shanghai Pujiang Program with Grant No. 20PJ1408100 (JW)

\begin{appendix}

\section{Detailed calculations for the 2D Dirac fermions}
\label{sec:2Ddetail}

We consider the Hamiltonian 
\begin{align}
 \hat{\mathcal{H}}
   =v[\xi(k_{x}-eA\sin q)\hat{\sigma}^{x}
        +(k_{y}+eA\cos q)\hat{\sigma}^{y}]
     +m\hat{\sigma}^{z}.
\nonumber
\end{align}
where $q=\Omega t$. 
It is expanded as to $q$ up to the second order and 
can be written in the form of 
Eq.~\eqref{eq:HamilGeneralOpSM} with 
\begin{align}
 \hat{A}=&m\hat{\sigma}^{z}
   +\xi v k_{x}\hat{\sigma}^{x}
   +v(k_{y}+eA)\hat{\sigma}^{y}\nonumber\\
 \hat{B}=&-\xi veA\hat{\sigma}^{x}\nonumber\\
 \hat{C}=&-veA\hat{\sigma}^{y}.\nonumber
\end{align}
Let us consider the tunneling at 
$k_{x}=0$, $k_{y}<0$ for $\Omega>0$ and 
$k_{x}=0$, $k_{y}>0$ for $\Omega<0$ 
(i.e., $k_{y}=-\mathrm{sgn}(\Omega)|\boldsymbol{k}|$) 
in the time interval of 
$-\pi/|\Omega|\leq t\leq \pi/|\Omega|$. 
Note that the sign of $A$ is the same as that of $\Omega$. 
The parameters in Eq.~\eqref{eq:Hamil_q2_SM} are given as 
\begin{align}
 m\to&\sqrt{v^{2}(-\mathrm{sgn}(\Omega)|\boldsymbol{k}|
   +eA)^{2}+m^{2}}\nonumber\\
 v\to&veA\nonumber\\
 \kappa_{\parallel}v^{2}
   \to&\frac{\xi mveA}
     {\sqrt{v^{2}(-\mathrm{sgn}(\Omega)|\boldsymbol{k}|+eA)^{2}
     +m^{2}}}.
\nonumber
\end{align}
Since $F=-\Omega$ and $E=A\Omega$, 
the tunneling probability for twisted Schwinger effect 
in 2D Dirac fermions is given as 
\begin{align}
\mathcal{P}_{\xi}(\boldsymbol{k})
   =\exp\Bigg[-\pi
     \frac{\displaystyle\Big(M-\frac{\xi\Omega m}{4M}\Big)^{2}}{veE}
     \Bigg],
\label{eq:P_MonoGraSM}
\end{align}
where we defined 
\begin{align}
 M=\sqrt{v^{2}(|\boldsymbol{k}|-eE/|\Omega|)^{2}+m^{2}}.
\nonumber
\end{align}
We investigate the total probability 
\begin{align}
 \mathcal{P}_{\xi}^{\mathrm{tot}}
   \equiv \Big(\frac{a}{2\pi}\Big)^{2}\int d\boldsymbol{k}
     \mathcal{P}_{\xi}(\boldsymbol{k})
   =\frac{a^{2}}{2\pi}\int_{0}^{\infty}dk k\mathcal{P}_{\xi}(k).
\nonumber
\end{align}
below focusing on the case of $\xi\Omega>0$.

\subsection{Low frequency region}
\label{sec:2Ddetail1}
When the laser frequency is smaller than double the gap 
$|\Omega|<4m$, 
$\mathcal{P}_{\xi}$ shows a peak at 
$|k|=eE/|\Omega|$ in the momentum space. 
Let expand Eq.~\eqref{eq:P_MonoGraSM} around $k=eE/|\Omega|$. 
We represent $k'=k-eE/|\Omega|$, 
and since $vk'\ll m$, we can approximate as 
$(m/M)^{2}=(1+v^{2}k'^{2}/m^{2})^{-1}\simeq 1-v^{2}k'^{2}/m^{2}$. 
Hence, in the low frequency region, 
\begin{align}
 \mathcal{P}_{\xi}(\boldsymbol{k})
   \simeq&\exp\Big[-\frac{\pi}{veE}
     \Big(m^{2}+v^{2}k'^{2}
     -\frac{\xi\Omega m}{2}
     +\frac{\Omega^{2}}{16}(1-v^{2}k'^{2}/m^{2})\Big)\Big]
\nonumber\\
   =&\exp\Big[-\frac{\pi}{veE}
     \Big\{v^{2}\Big(1-\frac{\Omega^{2}}{16m^{2}}\Big)k'^{2}
     +\Big(m-\frac{\xi\Omega}{4}\Big)^{2}\Big\}\Big],
\label{eq:pk_approx_lowfreqSM}
\end{align}
which is the normal distribution with the standard deviation 
$\sqrt{eE/(2\pi v)}(1-\Omega^{2}/(16m^{2}))^{-1/2}$. 
When 
$\sqrt{eE/(2\pi v)}(1-\Omega^{2}/(16m^{2}))^{-1/2}\ll eE/(|\Omega|)$, 
$\mathcal{P}_{\xi}^{\mathrm{tot}}$ can be calculated as 
\begin{align}
 \mathcal{P}_{\xi}^{\mathrm{tot}}
   \simeq&\exp\Big[-\frac{\pi}{veE}
     \Big\{v^{2}\Big(1-\frac{\Omega^{2}}{16m^{2}}\Big)k'^{2}
     +\Big(m-\frac{\xi\Omega}{4}\Big)^{2}\Big\}\Big]
\nonumber\\
   =&\frac{eEa^{2}}{2\pi|\Omega|}\sqrt{\frac{eE}{v}}
     \Big(1-\frac{\Omega^{2}}{16m^{2}}\Big)^{-1/2}
     \exp\Big[-\frac{\pi}{veE}\Big(m-\frac{\xi\Omega}{4}\Big)^{2}\Big].
\nonumber
\end{align}
Therefore the e-h production rate per unit of time and volume
is provided as 
\begin{align}
 \Gamma_{\xi}\equiv
   \frac{|\Omega|}{2\pi a^{2}}\mathcal{P}_{\xi}^{\mathrm{tot}}
   \simeq\frac{eE}{(2\pi)^{2}}\sqrt{\frac{eE}{v}}
     \exp\Big[-\frac{\pi}{veE}\Big(m-\frac{\xi\Omega}{4}\Big)^{2}\Big].
\label{eq:ptot_approx_lowfreqSM}
\end{align}

\subsection{High frequency region}
\label{sec:2Ddetail2}
In the high frequency region 
$ |\Omega|>4m$, 
$\mathcal{P}_{\xi}(\boldsymbol{k})$ have peaks 
at the perfect tunneling points 
\begin{align}
 k=\frac{eE}{|\Omega|}\pm
   \frac{1}{v}\sqrt{m(\xi\Omega/4-m)}
\nonumber
\end{align}
instead of $k=eE/|\Omega|$. 
In the case of strong electric field, however, 
the broadening of $\mathcal{P}_{\xi}(\boldsymbol{k})$ 
is much larger than the distance 
between the perfect tunneling points 
$\sqrt{eE/(2\pi v)}\gg\sqrt{m(\xi\Omega/4-m)}/v$ 
and the contribution to $\mathcal{P}_{\xi}^{\mathrm{tot}}$ 
mainly comes from 
$k<eE/|\Omega|-\sqrt{m(\xi\Omega/4-m)}/v$ 
and 
$k>eE/|\Omega|+\sqrt{m(\xi\Omega/4-m)}/v$, 
where the approximation Eq.~\eqref{eq:pk_approx_lowfreqSM} 
is still valid. 
Hence the e-h production rate per unit of time and volume 
is provided by Eq.~\eqref{eq:ptot_approx_lowfreqSM}. 

In the case of weak electric field, 
the contribution to $\mathcal{P}_{\xi}^{\mathrm{tot}}$ comes from 
the wavenumber around the perfect tunneling point 
$k=eE/(|\Omega|)+\sqrt{m(\xi\Omega/4-m)}/v$. 
By expanding $\mathcal{P}_{\xi}(\boldsymbol{k})$ 
around this wave number, i.e., 
$k=eE/|\Omega|+\sqrt{m(\xi\Omega/4-m)}/v+k'$, 
we obtain 
\begin{align}
 \mathcal{P}_{\xi}(\boldsymbol{k})
   \simeq& \exp\Big[
     -\xi\frac{4\pi}{veE\Omega m}
   \Big(vk'
     \sqrt{m(\xi\Omega-4m)}
     +v^{2}{k'}^{2}
   \Big)^{2}\Big]
\nonumber\\
 \simeq& \exp\Big[-\xi\frac{4\pi v}{eE\Omega}
   (\xi\Omega-4m){k'}^{2}\Big].
\end{align}
Thus, by noting 
$eE/|\Omega|\ll \sqrt{m(\xi\Omega/4-m)}/v$, 
the total production rate is given as 
\begin{align}
 \mathcal{P}_{\xi}^{\mathrm{tot}}
   \simeq \frac{a^{2}}{8\pi v}
     \sqrt{\xi\Omega m}\sqrt{\frac{eE}{v}}.
\end{align}
Thus the e-h production rate per unit of time and volume
is given as 
\begin{align}
 \Gamma_{\xi}
   =\frac{|\Omega|}{(4\pi)^{2}v}
     \sqrt{\xi\Omega m}\sqrt{\frac{eE}{v}}.
\end{align}

\section{Detailed calculations for the 3D Dirac fermions}
\label{sec:appendix3D}

In the same way as the 2D case, 
the tunneling probability for twisted Schwinger effect 
in 3D Dirac fermions is given as 
\begin{align}
\mathcal{P}_{\xi}(\tilde{\boldsymbol{q}})
   =\exp\Bigg[-\pi
     \frac{\displaystyle\Big(M
       -\frac{\xi\mathrm{sgn}(\Omega)\tilde{k}_{z}}{4M}\Big)^{2}}
       {\tilde{A}}
     \Bigg],
\end{align}
with 
\begin{align}
 M=\sqrt{(|\tilde{\boldsymbol{k}}|-\tilde{A})^{2}+\tilde{k}_{z}^{2}}.
\nonumber
\end{align}
Then the e-h production rate per unit time and volume 
for each chirality is given as 
\begin{align}
 \Gamma_{\xi}^{\mathrm{3D}}
   =\frac{|\Omega|}{2\pi a^{3}}
   \Big(\frac{a}{2\pi}\Big)^{3}
     \int d\boldsymbol{q}
     \mathcal{P}_{\xi}(\boldsymbol{q})
 =\frac{|\Omega|^{4}}{(2\pi)^{4}v^{3}}
   \int d\tilde{\boldsymbol{q}}
     \mathcal{P}_{\xi}(\tilde{\boldsymbol{q}}).
\label{eq:Ptot3D}
\end{align}
We can calculate the total and chiral 
e-h production rates as 
$\Gamma_{\mathrm{tot}}^{\mathrm{3D}}
=\Gamma_{+}^{\mathrm{3D}}+\Gamma_{-}^{\mathrm{3D}}$ and 
$\Gamma_{5}^{\mathrm{3D}}
=\Gamma_{+}^{\mathrm{3D}}-\Gamma_{-}^{\mathrm{3D}}$.

The main contribution to the production rates and currents 
come from the wavenumbers around the perfect tunneling points 
\begin{align}
 |\tilde{\boldsymbol{k}}|
   =\tilde{A}\pm\sqrt{\tilde{k}_{z}
     (\xi\mathrm{sgn}(\Omega)/4-\tilde{k}_{z})}
\quad
 (\xi\mathrm{sgn}(\Omega)/8-1/8
   \leq \tilde{k}_{z}
   \leq \xi\mathrm{sgn}(\Omega)/8+1/8).
\label{eq:kpeak3D}
\end{align}
Equation~\eqref{eq:kpeak3D} is rewritten as 
\begin{align}
 (|\tilde{\boldsymbol{k}}|-\tilde{A})^{2}
   +(\tilde{k}_{z}-\xi\mathrm{sgn}(\Omega)/8)^{2}
   =(1/8)^{2},
\nonumber
\end{align}
which forms a circle with the center 
$(\tilde{A},\xi\mathrm{sgn}(\Omega)/8)$ 
and the radius $1/8$ or a part of it 
in the $|\boldsymbol{k}|$-$k_{z}$ space.

\subsection{Weak field regime}

In the weak field regime $\tilde{A}\ll 1/8$, 
the contribution mainly comes from 
the positive sign branch of Eq.~\eqref{eq:kpeak3D} 
$ |\tilde{\boldsymbol{k}}|=\tilde{A}+
\sqrt{\tilde{k}_{z}(\xi\mathrm{sgn}(\Omega)/4-\tilde{k}_{z})}$. 
The tunneling probability can be approximated as 
\begin{align}
 \mathcal{P}_{\xi}(\tilde{\boldsymbol{q}})
   \simeq \exp\Big[-\xi\mathrm{sgn}(\Omega)\frac{4\pi}{\tilde{A}}
   (\xi\mathrm{sgn}(\Omega)-4\tilde{k}_{z})\tilde{k}'^{2}\Big],
\end{align}
where 
$\tilde{k}'=|\tilde{\boldsymbol{k}}|-\tilde{A}-\sqrt{\tilde{k}_{z}
(\xi\mathrm{sgn}(\Omega)/4-\tilde{k}_{z})}$. 
Then we can calculate the production rate Eq.~\eqref{eq:Ptot3D} as 
\begin{align}
 &\Gamma_{\xi}^{\mathrm{3D}}\bigg/
   \Big(\frac{|\Omega|^{4}}{v^{3}}\Big)
\nonumber\\
   &\simeq
   \frac{1}{(2\pi)^{4}}\int d\tilde{\boldsymbol{q}}
     \exp\Big[-\xi\mathrm{sgn}(\Omega)\frac{4\pi}{\tilde{A}}
   (\xi\mathrm{sgn}(\Omega)-4\tilde{k}_{z})\tilde{k}'^{2}\Big]
\nonumber\\
   &=\frac{1}{2(2\pi)^{3}}
     \int_{\xi\mathrm{sgn}(\Omega)/8-1/8}^{\xi\mathrm{sgn}(\Omega)/8+1/8}
     d\tilde{k}_{z}\Big(\tilde{A}+
     \sqrt{\tilde{k}_{z}(\xi\mathrm{sgn}(\Omega)/4-\tilde{k}_{z})}\Big)
     \sqrt{\frac{\tilde{A}}{\xi\mathrm{sgn}(\Omega)
       (\xi\mathrm{sgn}(\Omega)-4\tilde{k}_{z})}}
\nonumber\\
   &\simeq\frac{1}{4(2\pi)^{3}}
     \int_{\xi\mathrm{sgn}(\Omega)/8-1/8}^{\xi\mathrm{sgn}(\Omega)/8+1/8}
     d\tilde{k}_{z}\sqrt{\xi\mathrm{sgn}(\Omega)\tilde{A}\tilde{k}_{z}}
   =\frac{A^{1/2}}{6(4\pi)^{3}}.
\end{align}
Therefore 
\begin{align}
 \Gamma_{\mathrm{tot}}^{\mathrm{3D}}\bigg/
   \Big(\frac{|\Omega|^{4}}{v^{3}}\Big)
   =\frac{A^{1/2}}{3(4\pi)^{3}},\quad
 \Gamma_{5}^{\mathrm{3D}}\bigg/
   \Big(\frac{|\Omega|^{4}}{v^{3}}\Big)=0.
\end{align}

For the calculation of currents, noting that 
\begin{align}
 |\tilde{\boldsymbol{k}}|^{2}+\tilde{k}_{z}^{2}
   \simeq \Big(\tilde{A}
     +\sqrt{\tilde{k}_{z}(\xi\mathrm{sgn}(\Omega)/4-\tilde{k}_{z})}
     \Big)^{2}+\tilde{k}_{z}^{2}
   \simeq \xi\mathrm{sgn}(\Omega)\tilde{k}_{z}/4,
\nonumber
\end{align}
we can derive 
\begin{align}
 J_{\xi}^{z}\bigg/
   \Big(\frac{-e\tau|\Omega|^{4}}{v^{2}}\Big)
   \simeq&\frac{4}{(2\pi)^{4}}
     \int d\tilde{\boldsymbol{q}}
     \xi\mathrm{sgn}(\Omega)
     \sqrt{\xi\mathrm{sgn}(\Omega)\tilde{k}_{z}}
     \mathcal{P}_{\xi}(\tilde{\boldsymbol{q}})
\nonumber\\
   \simeq&\frac{1}{(2\pi)^{3}}
     \int_{\xi\mathrm{sgn}(\Omega)/8-1/8}^{\xi\mathrm{sgn}(\Omega)/8+1/8}
     d\tilde{k}_{z}\tilde{A}^{1/2}\tilde{k}_{z}
   =\xi\mathrm{sgn}(\Omega)\frac{\tilde{A}^{1/2}}{4(4\pi)^{3}}
\end{align}
Therefore
\begin{align}
 J^{z}\bigg/
   \Big(\frac{-e\tau|\Omega|^{4}}{v^{2}}\Big)=0,\quad
 J_{5}^{z}\bigg/
   \Big(\frac{-e\tau|\Omega|^{4}}{v^{2}}\Big)
   =\mathrm{sgn}(\Omega)\frac{\tilde{A}^{1/2}}{2(4\pi)^{3}}.
\end{align}

\subsection{Strong field regime}

In the strong field regime $\tilde{A}\gg 1/8$, 
the contribution mainly comes from 
both sign branches of Eq.~\eqref{eq:kpeak3D}. 
The tunneling probability can be approximated as 
\begin{align}
 \mathcal{P}_{\xi}(\tilde{\boldsymbol{q}})
   \simeq \exp\Big[-\frac{\pi}{\tilde{A}}
     \Big\{\tilde{k}'^{2}
     +\Big(\tilde{k}_{z}-\frac{\xi\mathrm{sgn}(\Omega)}{4}
     \Big)^{2}\Big\}\Big],
\end{align}
where 
$\tilde{k}'=|\tilde{\boldsymbol{k}}|-\tilde{A}$. 
Then we can calculate the production rate Eq.~\eqref{eq:Ptot3D} as 
\begin{align}
 \Gamma_{\xi}^{\mathrm{3D}}\bigg/
   \Big(\frac{|\Omega|^{4}}{v^{3}}\Big)
   \simeq &
   \frac{1}{(2\pi)^{4}}\int d\tilde{\boldsymbol{q}}
     \exp\Big[-\frac{\pi}{\tilde{A}}
     \Big\{\tilde{k}'^{2}
     +\Big(\tilde{k}_{z}-\frac{\xi\mathrm{sgn}(\Omega)}{4}
     \Big)^{2}\Big\}\Big]
\nonumber\\
   =&\frac{\tilde{A}^{3/2}}{(2\pi)^{3}}
     \int_{-\infty}^{\infty}
     d\tilde{k}_{z}\exp\Big[-\frac{\pi}{\tilde{A}}
     \Big(\tilde{k}_{z}-\frac{\xi\mathrm{sgn}(\Omega)}{4}
     \Big)^{2}\Big]
   =\frac{A^{2}}{(2\pi)^{3}}.
\end{align}
Therefore 
\begin{align}
 \Gamma_{\mathrm{tot}}^{\mathrm{3D}}\bigg/
   \Big(\frac{|\Omega|^{4}}{v^{3}}\Big)
   =\frac{2A^{2}}{(2\pi)^{3}},\quad
 \Gamma_{5}^{\mathrm{3D}}\bigg/
   \Big(\frac{|\Omega|^{4}}{v^{3}}\Big)=0.
\end{align}

For the calculation of currents, noting that 
$|\tilde{\boldsymbol{k}}|^{2}+\tilde{k}_{z}^{2}\simeq \tilde{A}^{2}$, 
we can derive 
\begin{align}
 J_{\xi}^{z}\bigg/
   \Big(\frac{-e\tau|\Omega|^{4}}{v^{2}}\Big)
   \simeq&\frac{2}{(2\pi)^{4}\tilde{A}}
     \int d\tilde{\boldsymbol{q}}
     \tilde{k}_{z}
     \mathcal{P}_{\xi}(\tilde{\boldsymbol{q}})
\nonumber\\
   \simeq&\frac{2\tilde{A}^{1/2}}{(2\pi)^{3}}
     \int_{-\infty}^{\infty}
     d\tilde{k}_{z}\tilde{k}_{z}
     \exp\Big[-\frac{\pi}{\tilde{A}}
     \Big(\tilde{k}_{z}-\frac{\xi\mathrm{sgn}(\Omega)}{4}
     \Big)^{2}\Big]
   =\xi\mathrm{sgn}(\Omega)\frac{\tilde{A}}{2(2\pi)^{3}}
\end{align}
Therefore
\begin{align}
 J^{z}\bigg/
   \Big(\frac{-e\tau|\Omega|^{4}}{v^{2}}\Big)=0,\quad
 J_{5}^{z}\bigg/
   \Big(\frac{-e\tau|\Omega|^{4}}{v^{2}}\Big)
   =\mathrm{sgn}(\Omega)\frac{\tilde{A}}{(2\pi)^{3}}.
\end{align}


\end{appendix}




\nolinenumbers

\end{document}